\documentclass[preprint]{elsarticle}

\pdfoutput=1

\usepackage{graphicx}
\usepackage{subfigure}
\RequirePackage{lineno} 
 
\usepackage{ifthen}
\usepackage{lettrine} 
\usepackage{boxedminipage}
\usepackage{hyperref}

\newboolean{BiblioVersion}
\setboolean{BiblioVersion}{true}

\usepackage{natbib} 

\usepackage{array}
\usepackage{color}
\usepackage{graphicx}
\usepackage{subfigure}
\usepackage{rotating}
\usepackage{amsmath}
\usepackage{amssymb}
\usepackage[T1]{fontenc}
\usepackage{epsfig}
\usepackage{multirow}
\usepackage{placeins}
\usepackage{longtable}
\usepackage{rotating}

\journal{Astroparticle Physics}

\begin{document}

\begin{frontmatter}

\author[label1]{Karl-Heinz Kampert}
\author[label2]{J\"org Kulbartz}
\author[label3,label4]{Luca Maccione}
\author[label1,label2]{Nils Nierstenhoefer}
\author[label2]{Peter Schiffer}
\author[label2]{G\"unter Sigl}
\author[label2]{Arjen Ren\'e van Vliet}
\address[label1]{University of Wuppertal, Gau\ss stra\ss e 20, 42119 Wuppertal, Germany}
\address[label2]{II. Institut f\"ur Theoretische Physik, Universit\"at Hamburg, Luruper Chaussee 149, 22761 Hamburg, Germany}
\address[label3]{Ludwig-Maximilians-Universit\"at, Arnold Sommerfeld Center,
Theresienstra{\ss}e 37, D-80333 M\"unchen}
\address[label4]{Max-Planck-Institut f\"ur Physik (Werner-Heisenberg-Institut),
F\"ohringer Ring 6, D-80805 M\"unchen}

\title{CRPropa 2.0 -- a Public Framework for Propagating High Energy Nuclei, Secondary Gamma Rays and Neutrinos}
 
\begin{abstract}
Version 2.0 of CRPropa\footnote{CRPropa is published under the 3rd version of the GNU General Public License (GPLv3). It is available, together with a detailed documentation of the code, at \url{https://crpropa.desy.de}.} is public software to model the extra-galactic propagation of ultra-high energy nuclei of atomic number $Z\leq 26$ through structured magnetic fields and ambient photon backgrounds taking into account all relevant particle interactions. CRPropa covers the energy range $6\times 10^{16} < E/\rm{eV} < A\times 10^{22}$ where $A$ is the nuclear mass number. CRPropa can also be used to track secondary $\gamma$-rays and neutrinos which allows the study of their link with the charged primary nuclei -- the so called multi-messenger connection. After a general introduction we present several sample applications of current interest concerning the physics of extragalactic ultra-high energy radiation.

\end{abstract}

\end{frontmatter}

\section{Introduction and Motivation}

Cosmic rays are ionized atomic nuclei reaching the Earth from outside the Solar System with energies that exceed $10^{20}~\rm eV$. Although ultra-high energy cosmic rays (UHECRs) were originally discovered in 1939, their sources and propagation mechanisms are still a subject of intense research. During the last decade significant progress has been made due to the advent of high quality and high statistics data from a new generation of large scale observatories. Observables of prime interest are the energy spectrum, mass composition and arrival direction of cosmic rays.
A flux-suppression in the energy spectrum above $E \sim 5 \cdot 10^{19}$\,eV has been observed by the HiRes and Pierre Auger Observatories \cite{2008PhRvL.101f1101A,Abbasi-08} and possibly also by the Telescope Array \cite{Matthews:2011zz} indicating either observation of the GZK-effect \cite{Greisen1964,ZatsepinKutzmin1966} or the limiting energy of the sources. 
Moreover, data of the Pierre Auger Observatory indicate that the arrival directions of the highest energy cosmic rays are correlated with the depth of nearby Active Galactic Nuclei (AGN) or more generally with the nearby extra-galactic matter distribution \cite{Auger2007}.
 
Additionally, measurements of the position of the shower maximum and its fluctuations by the Pierre Auger Collaboration suggest a significant fraction of heavy primaries above $10^{19}$\,eV \cite{Abraham:2010yv}.
However, HiRes \cite{2010PhRvL.104p1101A} and preliminary data of the Telescope Array \cite{Matthews:2011zz}), suggest a proton dominance in the same energy range.  Unfortunately, the limited number of observed events does not yet allow the extension of these measurements to the aforementioned cutoff energy. 
Independently of the mass composition, it is not uniquely settled yet if this flux depression is due to energy loss or maximum energy limitation of the sources. 

Clearly, a better understanding of all of these features and of the effects of cosmic ray propagation through the local Universe is mandatory.

UHECRs do not propagate freely in the Inter Galactic Medium (IGM). During their propagation they suffer from catastrophic energy losses in reactions with the intergalactic background light and are deflected by poorly known magnetic fields. Thus, the effects of propagation alter the cosmic ray spectrum and composition injected by sources in the IGM and form the features detected by UHECR observatories. In order to establish the origin of UHECRs, it is of prime interest to quantitatively understand the imprint of the propagation and to disentangle it from the properties of the cosmic rays at their sources. In this respect, it is essential to compare the measured UHECR spectrum, composition and anisotropy with model predictions. This requires extensive simulations of the propagation of UHE nuclei and their secondaries within a given scenario. In particular, the observation that UHECRs may consist of a significant fraction of heavy nuclei challenges UHECR model predictions and propagation simulations. Indeed, compared to the case of ultra-high energy (UHE) nucleons, the propagation of nuclei leads to larger deflections in cosmic magnetic fields and additional particle interactions have to be taken into account, namely, photodisintegration and nuclear decay. 

To provide the community with a versatile simulation tool we present in this paper a publicly available Monte Carlo code called CRPropa~2.0 which allows one to simulate the propagation of UHE nuclei in realistic one- (1D) and three-dimensional (3D) scenarios taking into account all relevant particle interactions and magnetic deflections. To this end, we extended the former version 1.4 of CRPropa, which was restricted to nucleon primaries, to the propagation of UHE nuclei. CRPropa~1.4 provided an excellent basis for this effort as many of its features could be carried over to the case of UHE nuclei propagation. In the present paper, which accompanies the public release of CRPropa 2.0, the underlying physical and numerical frameworks of the implementation of nuclei propagation are introduced. For technical details the reader is referred to the documentation distributed along with this framework.

This paper is organized as follows: Section \ref{sct:CRPRopa1} starts with a short introduction of the publicly available previous CRPropa\,1.4. The extensions which were implemented for nuclei interactions in CRPropa\,2.0 are the subject of section \ref{sct:CRPropa:Interactions}. Section~\ref{Sec:PropAlgo} describes the general propagation algorithm and in section~\ref{sct:Applications} example applications of nuclei propagation with CRPropa are presented. We present a short summary and an outlook in Section~6.

Unless stated otherwise, we use natural units $\hbar = c = 1$ throughout this paper.

\section{Inherited features from CRPropa~1.4}\label{sct:CRPRopa1}

The previous version 1.4 of CRPropa is a simulation tool aimed at studying the propagation of neutrons and protons in the intergalactic medium. It provides a one-dimensional (1D) and a three-dimensional (3D) mode. In 3D mode, magnetic field- and source distributions can be defined on a 3D grid. This allows one to perform simulations in realistic source scenarios with a highly structured magnetic field configuration as provided by, e.g., cosmological simulations. In 1D mode, magnetic fields can be specified as a function of the distance to the observer, but their effects are obviously restricted to energy losses of $e^+e^-$ pairs due to synchrotron radiation within electromagnetic cascades. Furthermore, it is possible to specify the cosmological and the source evolution as well as the redshift scaling of the background light intensity in 1D simulations. All important interactions with the cosmic infrared (IRB) and microwave (CMB) background light are included, namely, production of electron-positron pairs, photopion production and neutron decay. Additionally, CRPropa allows for tracking and propagating secondary $\gamma-$rays, $e^{+}e^{-}$ pairs and neutrinos. A module \cite{Lee:1996fp} is included that solves the one-dimensional transport equations for electromagnetic cascades that are initiated by electrons, positrons or photons taking into account single, double and triple pair production as well as up-scattering of low energy background photons by inverse Compton scattering. Synchrotron radiation along the line of sight can also be simulated. 

Technically, CRPropa is a stand alone object-oriented C++ software package. It reads an input file which specifies technical parameters as well as details of the simulated ``Universe'' such as source positions and magnetic fields. The CRPropa simulations for a given scenario generate output files of either detected events or full UHECR trajectories.

\section{Modeling nuclei interactions in CRPropa~2.0 \label{sct:CRPropa:Interactions}}
Similar to the case of protons, nuclei carry charge and suffer energy losses by electron-positron pair production in ambient photon fields. This can occur when photon energies boosted into the rest frame of the nucleus are of the order of $\epsilon^\prime\sim 1~\rm{MeV}$. For photon energies at or above the nuclear binding energy $\epsilon^\prime \gtrsim 8-9~\rm{MeV}$, nucleons and light nuclei can be stripped off the nucleus (photodisintegration). Finally, at photon energies exceeding $\epsilon^\prime \sim$145~MeV the quark structure of free or bound nucleons can be excited to produce mesons (photopion production). In these reactions the nucleus can be disrupted and unstable elements be produced. Hence, nuclear decay has to be taken into account as well. 

In CRPropa\,2.0 a nucleus with energy $E$ and mass number $A$ is considered a superposition of $A$ nucleons with energy $E/A$. Thus, if one or several nucleons are stripped off, the initial energy $E$ will be distributed among the outgoing nucleons and nuclei. The ultra-relativistic limit $\beta\to1$ (with $\beta=v/c$ of the nucleus) is used in CRPropa such that all nuclear products are assigned the same velocity vector as the initial particle. This corresponds to a Lorentz factor $\Gamma\simeq10^8 \cdot \left(E/10^{17}\,{\rm eV}\right)\cdot A^{-1}$ and a forward collimation within an angle $\simeq1/\Gamma$.


\subsection{Photodisintegration}
Photodisintegration of nuclei has no analogy for free nucleons. Thus, implementing this new interaction process is mandatory to allow for propagation of nuclei within CRPropa. There are many competing photodisintegration processes of different cross sections which need to be accounted for along the path of the nucleus in the photon field. Thus, it is important to efficiently describe the specific photodisintegration pattern of each propagated nucleus in CRPropa. 

The effects of the propagation of UHE nuclei have first been studied by Puget, Stecker and Bredekamp (PSB) \cite{Puget:1976nz}. The approach to model the photodisintegration process chosen in CRPropa 2.0 is similar to what was more recently discussed in Ref.~\cite{2005APh....23..191K}. Further details on the photodisintegration within CRPropa\,2.0 can be found in Ref.~\cite{NierstenhoeferPHD}.
As target photon fields we shall consider the CMB and IRB, for which we adopt the more recent parametrization developed in \cite{Kneiske}.

\subsubsection{The Photonuclear Cross Sections \label{sct:CRPropa:ThePDXS}}

We use the publicly available TALYS framework, version 1.0~\cite{TALYS} to compute photodisintegration cross sections. The nuclear models therein are reliable for mass numbers $A\geq 12$.\footnote{Nevertheless, TALYS provides results for nuclei with $A>5$ and $N>2$ where $N$ is the number of neutrons.}  Thus, additional photodisintegration cross sections for light nuclei have to added in the modeling. In CRPropa\,2.0 TALYS was applied to 287 isotopes up to iron ($Z=26$)\footnote{One could easily extend the framework to nuclei heavier than iron. However, this is not done in CRPropa\,2.0 because there is no significant contribution of elements with $Z>26$ in the galactic cosmic ray composition.} employing nuclear models and settings as suggested in Ref.~\cite{2005APh....23..191K}. The list of isotopes for which the cross sections were calculated was generated using data from Ref.~\cite{NuDatHP}. It is assumed that excited nuclei will immediately return to their ground state. Hence, only nuclei in their ground states are considered when calculating the cross sections\footnote{In this way we also neglect the photons created in nuclear de-excitations. Note that these photons might contribute to the overall flux of photons at energies ($E\lesssim 0.1-1~\rm EeV$).}. All cross sections were calculated for photon energies $1\,{\rm keV}\leq\epsilon^\prime\leq\,250\,$MeV in the rest frame of the nucleus and stored in 500 bins of energy. Knocked out neutrons ($n$), protons ($p$), deuterium ($d$), tritium ($t$), helium-3 ($^{3}\rm{He}$) and helium-4 ($\alpha$) nuclei and combinations thereof are considered by TALYS. The corresponding reaction channels are called \textit{exclusive channels}. 
In the mass range of target nuclei where TALYS cannot be employed reliably, we use instead other prescriptions, as follows:
\begin{itemize}
\item 
$^9$Be, $^4$He, $^3$He, $t$ and $d$ as given in Ref.~\cite{RachenPHD}. 
\item
A parametrization of the total photonuclear cross section as function of the mass number $A$ is used for $^8$Li, $^9$Li, $^7$Be, $^{10}$Be, $^{11}$Be, $^{8}$B, $^{10}$B, $^{11}$B, $^{9}$C, $^{10}$C, $^{11}$C as described in \cite{2002EPJA...14..377K}. In these cases the loss of one proton (neutron) is assumed if the neutron number $N<Z$ ($N>Z$). For $N=Z$, the loss of one neutron or proton is modeled with equal probability.
\item
For $^7$Li we use experimental data from Ref.~\cite{B_CDFELI2__1986,J_IZV_27_11_1412_63} and instead of using a parametrization we interpolate linearly between the measured data points. 
\end{itemize} 
In total 78449 exclusive channels are taken into account.

The TALYS output in general agrees reasonably well with available measured data and only in rare cases differs up to a factor of 2 for the integrated (total absorption) cross sections \cite{2005APh....23..191K}.

Alternatively, for comparison, the widely-used photodisintegration cross section estimates developed by Puget, Stecker and Bredekamp \cite{Puget:1976nz} can be used.

In the PSB case a reduced reaction network is implemented involving one nucleus for each atomic mass number $A$ up to $^{56}$Fe. Herein, the cross sections for one- and two-nucleon dissociation in the photon energy interval $\epsilon_{min}^\prime\leq\epsilon^\prime\leq\,30\,$MeV are parametrized by a Gaussian approximation. Different from \cite{Puget:1976nz}, the channel-dependent threshold energies $\epsilon_{min}^\prime$ proposed in \cite{Stecker:1998ib} are used. In the photon energy interval $30\,\leq\epsilon^\prime\leq\,150\,$MeV the cross section is assumed to be constant. A comparison of results obtained with the TALYS and PSB cross sections is given in section \ref{sct:Applications:PSB}.

\subsubsection{Mean Free Path Calculations and Channel Thinning. \label{sct:CRPropa:MathematicalFormulation}}
Once the photodisintegration cross section in the nucleus rest frame $\sigma(\epsilon^\prime)$ is given (cf.\ Sec.\,\ref{sct:CRPropa:ThePDXS}), the energy weighted average cross section
\begin{equation}
  \bar\sigma(\epsilon_{\max}^\prime)
   = \frac{2}{(\epsilon_{\max}^\prime)^2}\,
  \int_{0}^{\epsilon_{\max}^\prime} \epsilon^\prime\sigma (\epsilon^\prime)\,d\epsilon^\prime.
  \label{eq:CRPRopa:MFPInPhotonDensity}
\end{equation}
is tabulated as a function of $\epsilon^\prime_{\max}=2\Gamma \epsilon$.

For a given  $\bar\sigma(\epsilon_{max}^\prime)$, for each isotope the mean free path $\lambda(\Gamma)$ can be calculated as a function of the Lorentz factor according to Ref.~\cite{Puget:1976nz},
\begin{equation}
  \lambda(\Gamma)^{-1} =  
  \int_{\epsilon_{\min}}^{ \epsilon_{\max} } n(\epsilon , z)\,
  \bar\sigma( \epsilon_{\max}^\prime=2\Gamma\epsilon)\,
  d\epsilon\,,
  \label{AttLengthInt}
\end{equation}
where $n(\epsilon,z)$ is the number density of the isotropic low energy photons per energy interval and volume. Variation due to cosmological redshift (cf.\ Sec.\,\ref{sct:PhotonFields}) is accounted for. For performance reasons, $\lambda(E)$ is tabulated as a function of energy. 
The values of the integration limits $\epsilon_{\min}$, $\epsilon_{\max}$ in Eq.~(\ref{eq:CRPRopa:MFPInPhotonDensity}) and~(\ref{AttLengthInt}) are listed in Tab.~\ref{tab:PDMFPCalcParamVal}. 

\begin{table}
\begin{center}
\caption{Values of the parameters used in Eq.~(\ref{eq:CRPRopa:MFPInPhotonDensity}) and~(\ref{AttLengthInt}) to create the mean free path tables for photodisintegration in CRPropa.}
 \label{tab:PDMFPCalcParamVal}
\begin{tabular}{|c|c|c|}
\hline
  parameter & CMB & IRB \\ \hline \hline  
  $\epsilon_{\min}$ (GeV)   & $4.00\times10^{-19}$ & $10^{-12}$\\ \hline
  $\epsilon_{\max}$ (GeV)  & $10^{-11}$ &  $10^{-7}$\\ \hline
\end{tabular}
\end{center}
\end{table}

\begin{figure}
   \centering
     
       \includegraphics[width=1.\textwidth]
                       {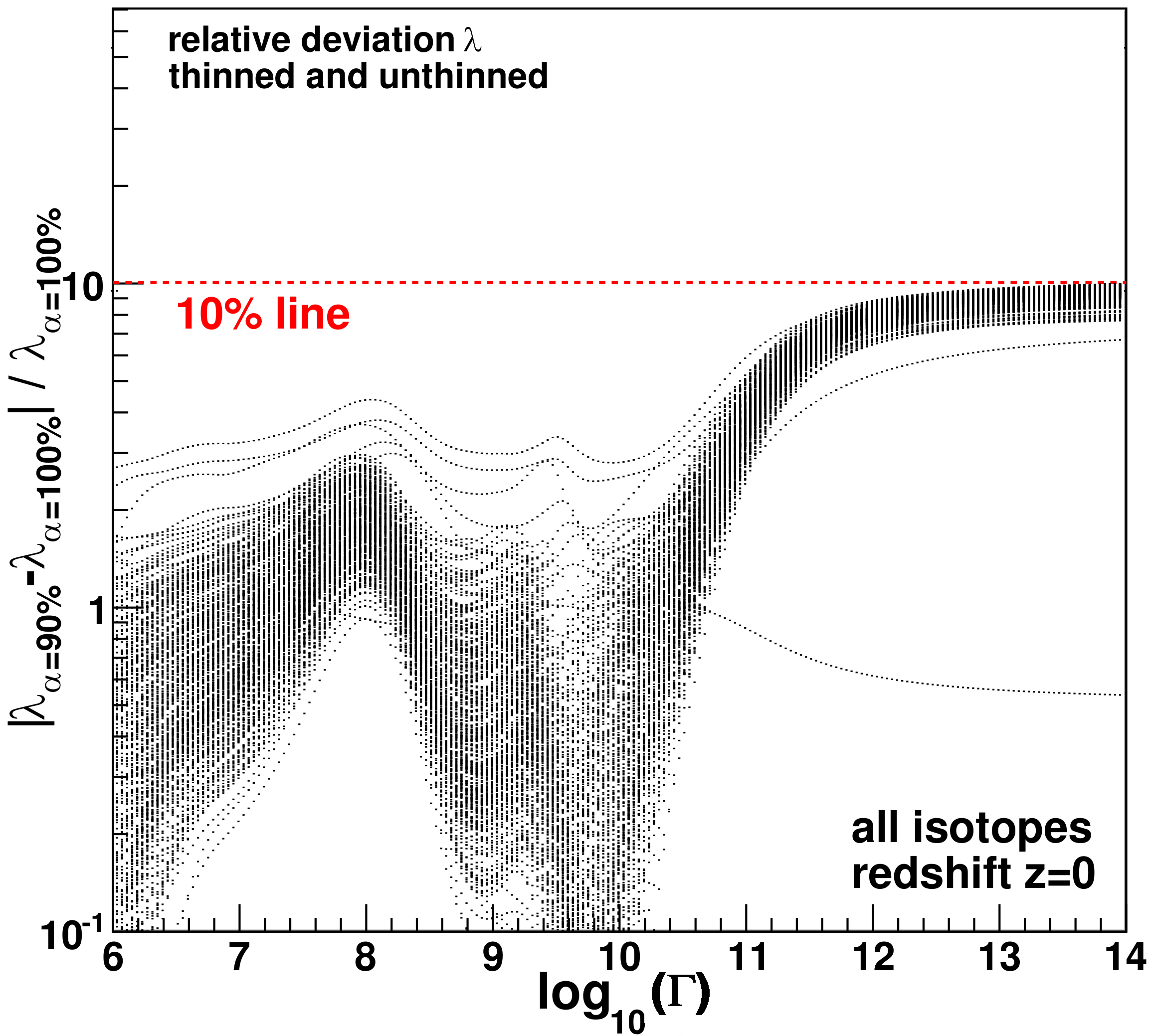}
                       
     \caption{The relative deviation of the total mean free path $\lambda$ in photodisintegration reactions in the CMB and IRB for the thinned ($\alpha=90\%$) and un-thinned case is given for all 287 isotopes (redshift $z=0$). 
       \label{CRPropa:ThnOpt} 
     }
 \end{figure}
 
As the Monte Carlo rapidly slows down with increasing number of exclusive channels to be sampled, a thinning procedure was implemented: For each isotope, we include only the channels with the $n$ largest interaction rates out of the $N$ available exclusive channels, such that the sum $ \sum_i^n \lambda^{-1}_{i}/\lambda^{-1}_{tot}>\alpha$ in at least one energy bin. Here, $\alpha$ is the thinning factor and the $\lambda^{-1}_{i}$ are summed up in decreasing order. Furthermore, $\lambda^{-1}_{\rm{tot}}=\sum_i^N \lambda_{i}^{-1}$ is calculated for each isotope. In this way a $\alpha=90\%$ channel-thinning reduces the number of photodisintegration channels to be tracked from 78449 to 6440. The thinning procedure leads to a systematic overestimation of the mean free path in the order of 1\% for Lorentz factors of the UHECR below $10^{10}$. This deviation goes up to 10\% for Lorentz factors above $10^{12}$ (see Fig.~\ref{CRPropa:ThnOpt}).

\subsection{Photopion Production}

UHE nucleons can produce secondary mesons in interactions with low energy photon backgrounds. The most important reaction of this type is the production of pions in reactions of UHE protons with the CMB, which leads to the well known GZK cut-off~\cite{Greisen1964,ZatsepinKutzmin1966} at an energy of about $E_{\rm GZK}=5\cdot 10^{19}$~eV. 

Nuclei can also produce mesons, albeit with a higher threshold energy of $E_{\rm th}\approx A \times E_{\rm GZK}$. This is due to the fact that to good approximation the center of mass system (CMS) coincides with the rest frame of the nucleus. The threshold energy, therefore, does not depend on the total energy of the nucleus $E$, but on the Lorentz factor $\Gamma\propto E/A$. Pion production is thus only relevant for extremely energetic nuclei. However, it is important to be included to properly account for production of secondary UHE photons and neutrinos, as well as for the propagation of secondary and primary protons and neutrons. Effectively this process leads to an energy scaling of the mean free path $\lambda_{A,Z}$ for photopion production of nuclei. The mean free path for pion production on the constituent protons $\lambda_p$ and neutrons $\lambda_n$ can thus we written as 
\begin{equation}    
  \lambda_{A,Z}^{-1}(E)=Z\,\lambda_{\rm{p}}^{-1}\left(\frac{E}{A}\right)\,+\,(A-Z)\,\lambda_{\rm{n}}^{-1}\left(\frac{E}{A}\right).
\label{sct:CRPropa:PiPLambda}
\end{equation}

In CRPropa 2.0 we use Eq.~(\ref{sct:CRPropa:PiPLambda}) to reduce the mean free path for pion production by nuclei to the one for nucleons which in turn is modeled by the SOPHIA package which was already used in CRPropa 1.3. This approximation is sufficient for our purposes because the pion mass $m_\pi\approx 140$~MeV is much larger than the binding energy per nucleon $E_b/A\lesssim8$~MeV so that, above the threshold for pion production, the nuclear binding energy can be neglected and the nucleus be treated as a collection of free nucleons. Following this argument, we split the reaction into four parts, and calculate reactions of protons and neutrons on the CMB and on the IRB separately. The nucleus cross section $\sigma_{A, Z}$ for these four channels is then given by the cross sections of protons or neutrons times the number of the respective nucleons in the nucleus. The mean free path of the reaction was then obtained by folding this cross section with the respective photon background.

If a reaction takes place, we treat it as a reaction of a free nucleon. The interacting nucleon will suffer energy loss and be stripped off the nucleus. The disintegrated nucleon will then be propagated individually, while the produced meson will decay leading to secondary leptons, photons or neutrinos which are then propagated using the corresponding modules of CRPropa. Both the decay of the meson as well as the energy loss of the primary nucleon are calculated by using the SOPHIA package~\cite{2000CoPhC.124..290M}. 

\subsection{Pair production \label{sct:PaP}}
Another interaction relevant for the propagation of UHE protons and nuclei is the creation of electron positron pairs in the low energy photon backgrounds. Both photomeson- and pair-production are less important in terms of energy loss of the primary nuclei which is dominated by photodisintegration (c.f.\  Fig.~\ref{CRPropa:AllReactionsInPlot}). Pair production is, however, the most important reaction for the creation of secondary photons in the TeV range. The mean free path for pair production is short, but the energy loss in each individual reaction is small. Thus, we treat pair production as a continuous energy loss which for interactions with the CMB can be parametrized by~\cite{Puget:1976nz}
\begin{equation}
-\frac{dE_{A,Z}^{e^{+}e^{-}}}{dt}=3\alpha_{\rm em}\sigma_{T}h^{-3}Z^2(m_{e}c^2k_{\rm B}T)f\left(\Gamma\right)\,.
\label{eq:ppbb}
\end{equation}
Here, $\sigma_T$ is the Thomson cross section, $m_e$ and $m_p$ are the electron and proton rest masses, respectively, $\alpha_{\rm em}$ is the fine structure constant, $T$ is the temperature of the CMB, and $f(\Gamma)$ is a function which depends only on the Lorentz factor $\Gamma$ and was parametrized by Blumenthal \cite{PhysRevD.1.1596}.
One can therefore express the energy loss length $l_{A,Z}=E \left(dE_{A,Z}^{e^{+}e^{-}}/dt\right)^{-1}$ for nuclei in terms of the energy loss length for protons $l_p=E\left(dE_{1,1}^{e^{+}e^{-}}/dt\right)^{-1}$, according to
\begin{equation}
l_{A,Z}\left(\Gamma\right)=\frac{Z^2}{A}l_p\left(\Gamma\right)\,.
\label{eq:ScalingRelPairP}
\end{equation}
This scaling relation holds for arbitrary target photon backgrounds since the prefactor can be traced back to the scaling of the cross section and to the definition of the energy loss length. Eq.~(\ref{eq:ScalingRelPairP}) is used in CRPropa 2.0 to generalize the pair production loss rates from protons to nuclei, which in practice is obtained by integration over the corresponding secondary spectra as parametrized by \cite{2008PhRvD.78c4013K}.

The energy loss for $e^+e^-$ pair production is calculated after each timestep $\Delta t$ and is therefore taken into account at discrete positions and times. CRPropa can also propagate secondary electromagnetic cascades initiated by the $e^+e^-$ pairs or by the $\gamma$-rays resulting from $\pi^0$-decay. For the injected secondary spectra we use the parametrization given by Kelner and Aharonian~\cite{2008PhRvD.78c4013K}. It should be noted that, in particular, close to an observer a large time step can degrade the accuracy of the propagated spectra, due to the discrete injection of the electromagnetic cascade. We refer the reader to the CRPropa 2.0 manual for details.

\subsection{Nuclear Decay}

For the propagation of UHECRs, nuclear decay is relevant, if unstable particles are produced by photodisintegration or photopion production. On the one hand, nuclear decay can change both the nucleus type and its energy, while on the other hand it technically ensures that unstable nuclei decay back to stable nuclei whose photodisintegration cross sections are known.

In CRPropa, decays are modeled as a combination of $\alpha$, $\beta^\pm$ decays and dripping of single nucleons ($p, n$). The decay length of a nucleus is given by its life time $\tau$ and the Lorentz factor $\Gamma$ to be

\begin{equation}\label{eq:decay_length}
\lambda_{\rm decay}=\Gamma\tau.
\end{equation}

In case of $p, n$ dripping and $\alpha$ decay, the decay products are assumed to inherit the Lorentz factor $\Gamma$ from the parent nucleus. This assumption is justified since the binding energy per nucleon is small compared to the masses of the decay products. The energy of all produced nuclei are, therefore, simply given by
\begin{align}
E_{A', Z'}=\Gamma m_{A', Z'}.
\label{eq:DecayELoss}
\end{align}

In case of $\beta^\pm$ decay we also use Eq.~(\ref{eq:DecayELoss}) and the momenta of $e^\pm$ and the neutrino are calculated from a three body decay (see e.g.\ \cite{Basdevant:2005in}) and are then boosted to the simulation frame.

\begin{figure}[tbp]
\begin{center}

\includegraphics[width=\textwidth]{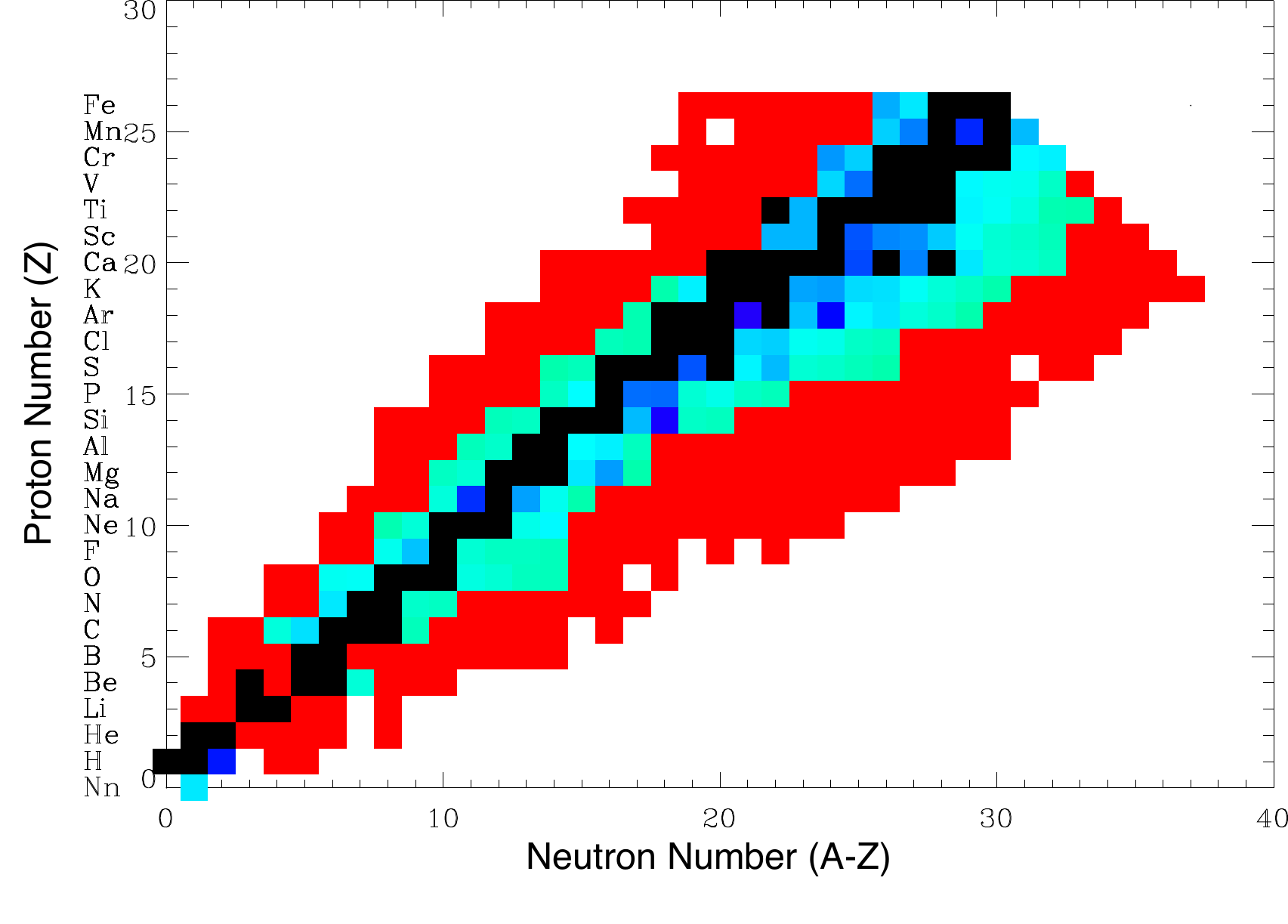}
\end{center}
\caption{Nuclear maps of the internal database. The life time is color-coded, where red corresponds to decay times less than $1$~s corresponding to a decay length $\le 0.1$~Mpc at a Lorentz factor of $\Gamma=10^{13}$. Black denotes effectively stable nuclei with a lifetime larger than $4\cdot10^{10}$~s. This corresponds to a decay length larger than $1/H_0$ at Lorentz factors of $\Gamma>10^7$. Nuclei between these two extremes are shown in blue, where darker shades correspond to longer decay times.}
\label{fig:NucMap}
\end{figure}

In CRPRopa the decay channels of the different nuclei as well as their decay constants at rest are stored in an internal database. It is based on the NuDat2 database \cite{NuDatHP} and contains 434 different nuclides with mass number $A \le 56$ and charge $Z\le 26$. It should be noted that UHECRs, unlike the isotopes in the NuDat2 database, are fully ionized. This means that electron capture (EC) is not possible for UHECRs and the $\beta^+$ decays have to be calculated from the EC rates given in the NuDat2 database. Up to the squared matrix elements which are the same for EC and $\beta^+$ decay, the rates $\tau^{-1}_{\rm EC}$ and $\tau^{-1}_{\beta^+}$ for these two processes are just proportional to the available phase space of the final state products.  If $\Delta m\equiv m_{A,Z}-m_{A',Z'}$ is the mass difference of the fully ionized nuclei and, since the nuclear recoil energy and the electron binding energy can be neglected, the kinetic energy of the final state leptons are given by the so-called Q-factors $Q_{\beta^+}=\Delta m-m_e$ and $Q_{\rm EC}=E_\nu=\Delta m+m_e=Q_{\beta^+}+2m_e$ with $E_\nu$ the neutrino energy. Therefore, $\tau^{-1}_{\rm EC}\propto|\psi_e(0)|^2Q_{\rm EC}^2/\pi$ with $|\psi_e(0)|^2=(Z/a_0)^3$ the normalized density of the electron wave function at the nucleus and $\tau^{-1}_{\beta^+}\propto(2/\pi^3)\int_{m_e}^{\Delta m}dE\,f(E)$ with $f(E)=E\sqrt{E^2-m_e^2}(\Delta m-E)^2$ the standard $\beta^+$ decay differential phase space density per total positron energy $E$ (including rest mass)~\cite{Basdevant:2005in}. Thus we get
\begin{align}
\frac{\tau_{\rm EC}}{\tau_{\beta^+}}= \frac{2}{\pi^2}\left(\frac{a_0}{Z}\right)^3\frac{\int_{m_e}^{\Delta m} f(E)\,dE}{(\Delta m+m_e)^2}\,,
\end{align}
where everything has been expressed in terms of the bare nucleus mass difference $\Delta m$. For all channels involving $\beta^{+}$ decay (including compound channels), we compute $\tau_{\beta^{+}}$ from the lifetime of the dressed nucleus given in NuDat2 $\tau$ by multiplying with $\tau_{\beta^+} / \tau = 1 + \tau_{\beta^+}/ \tau_{\rm EC}$. The resulting lifetimes $\tau_{\beta^{+}}$ of all isotopes in the database are shown in Fig.~\ref{fig:NucMap}. In Tab.~\ref{tab:ecCorrection} we list the isotopes for which $\tau_{\beta^{+}}$ deviates most from $\tau$.

In Fig.~\ref{CRPropa:AllReactionsInPlot} the decay length  is shown in comparison to other energy loss processes using the example of $^{47}$Ca.

\begin{table}[tbp]
\caption{All isotopes with $\tau_{\beta^{+}}$ differing from $\tau$ by more than $1\%$. Note that in the case of $^{36}Cl$ the very low branching ratio of $1.6\%$ does not lead to a strong modification of the propagation properties. In addition, there are only $3$ isotopes where $\tau_{\beta^{+}}$ is more than $10\%$ larger than $\tau$, $^{45}Ti$, $^{48}Cr$ and $^{54}Mn$, since $^{36}Cl$ and $^{40}K$ decay mostly by $\beta^-$ decay. Not included in the table are $10$ isotopes which only decay via electron capture due to their low mass difference $\Delta m < m_e$.}
\begin{center}
\begin{tabular}{|c|c|c|}
\hline
Isotope & $\tau_{\beta^+}/\tau$ & $\Delta m / {\rm keV}$\\
\hline
\hline
$^{18}F$  & $1.03$& 1145\\
$^{36}Cl$ & $39.8$& 631\\
$^{40}K$  & $1.56$& 993\\
$^{43}Sc$ & $1.05$& 1710\\
$^{45}Ti$ & $1.19$& 1551\\
$^{47}V$  & $1.01$& 2420\\
$^{50}V$  & $1.06$& 1696\\
$^{48}Cr$ & $1.48$& 1144\\
$^{49}Cr$ & $1.03$& 2117\\
$^{51}Mn$ & $1.01$& 2697\\
$^{54}Mn$ & $4.41$&  866\\
\hline
\end{tabular}
\end{center}
\label{tab:ecCorrection}
\end{table}

\subsection{Photon fields and cosmological evolution\label{sct:PhotonFields}}
The implementation of photodisintegration and pion production in CRPropa\,2.0 is based on tabulated mean free path data calculated with the photon density $n(\epsilon, z=0)$ at a redshift $z=0$ (Sec.\,\ref{sct:CRPropa:MathematicalFormulation}). As the photon density $n(\epsilon, z)$ evolves as a function of $z$, $\lambda=\lambda[\Gamma, z]$ is effectively altered, too. To model this change of $\lambda$ as function of $z$, a scaling function $s(z)$ is used. It approximately relates $\lambda[\Gamma, z]$ at redshift $z$ with the available tabulated data of $\lambda[\Gamma\,\,(1+z), z=0]$ at redshift $z=0$. For this scaling function $s(z)$, it is assumed that the normalized spectral shape of the photon field $n(\epsilon,z)$ does not change as function of $z$ in the comoving cosmological frame. In this approach the evolution of the photon number density $n(\epsilon, z)$ can be absorbed by a separated evolution factor $e(z)$
\begin{eqnarray}
n\left(\epsilon, z\right)\,\,=\,\,\left(1+z\right)^2\,\,n\left(\frac{\epsilon}{1+z},0\right)\,\,e\left(z\right). 
\label{eq:SeparationOfPhotonDensity}
\end{eqnarray}
In the approximation of a redshift independent spectral shape of $n(\epsilon, z)$, the evolution factor is defined by 

\begin{equation}
e(z)=\left\{
\begin{array}{l r}
 1 & \rm{CMB}\\
 \frac{\int_{\epsilon_{i}}^{\infty}\,\,n(\epsilon, z)\,\,d\epsilon}{\int_{\epsilon_{i}}^{\infty}\,\,n(\epsilon, 0)\,\,d\epsilon} & \rm{IRB} 
\end{array}
\right.
\end{equation}
Here, $\epsilon_{i}$ is the intersection energy of the CMB and IRB photon number densities $n_{\textrm{CMB}}(\epsilon_{i}, z)=n_{\textrm{IRB}}(\epsilon_{i}, z)$ in the comoving frame.

Substitution of Eq.~(\ref{eq:SeparationOfPhotonDensity}) in Eq.~(\ref{AttLengthInt}) gives the scaling relation for the mean free path  
\begin{eqnarray}
\lambda^{-1}[\Gamma, z]\,\,=\,\,(1+z)^3\,\,e(z)\,\,\lambda^{-1}[\Gamma\,\,(1+z), z=0]
\end{eqnarray}
from which one can find $s(z)=(1+z)^3\,\,e(z)$. It should be noted that this result is valid under the assumption that the spectral shape of the IRB does not depend on redshift. This approximation does not exactly hold for the IRB, due to energy injection in the IGM from galaxy formation. Nevertheless this approximation provides a model for the redshift evolution of the IRB which is of importance for the production of secondary neutrinos \cite{Kotera:2010yn}.
\section{Propagation Algorithm and Monte Carlo Approach}\label{Sec:PropAlgo}

To handle the widely ranging reaction rates of UHE nuclei, a new propagation algorithm has been implemented in CRPropa\,2.0. The main assumption is that the mean free paths $\lambda$ are approximately constant during a time step. As $\lambda=\lambda(E)$ is in general a function of the UHECR energy $E$, the numerical step size has to be small enough to ensure that no significant energy loss occurs.

\begin{figure}[tbp]
  \centering
    
      \includegraphics[width=1.\textwidth]
                      {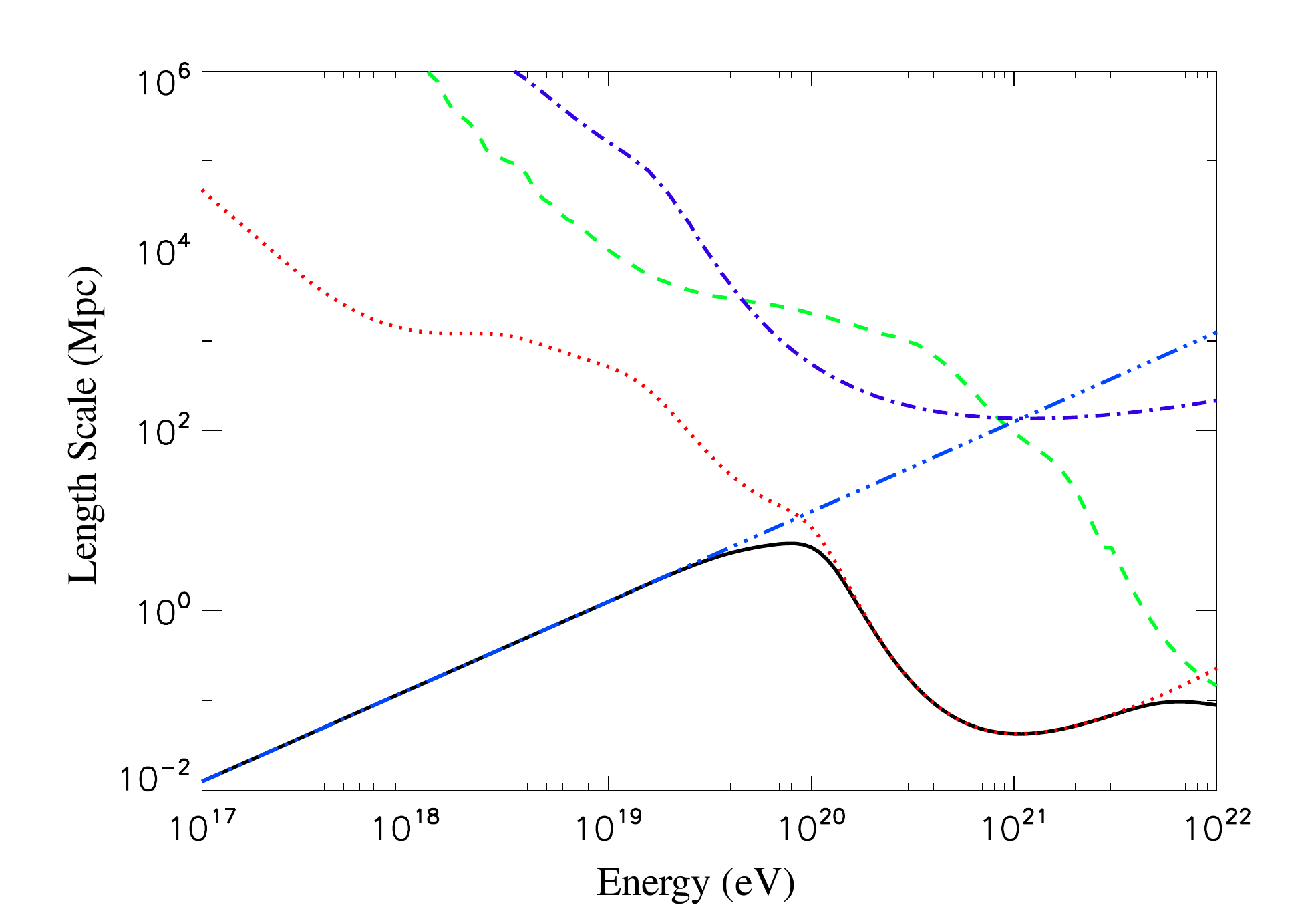}

\caption{Length scales for all interactions of $^{47}Ca$ as used by CRPropa. In the plot the dark blue (dash-dotted) line denotes the energy loss length of pair production. The red (dotted) and green (dashed) lines are the photodisintegration and pionproduction mean free paths respectively and the blue (dash-triple-dotted) line shows the decay length. The black (solid) line is the total mean free path of the catastrophic energy losses as used in the propagation algorithm. $^{47}Ca$ was chosen because the half life time of $4.5$~days corresponds to a decay length comparable to the other interactions. \label{CRPropa:AllReactionsInPlot}}

\end{figure}

The algorithm works as follows. 
Given the mean free path $\lambda_{i}$ for a given interaction channel of a given nucleus, where $i$ runs over all $N$ possible interaction and decay channels,
\begin{enumerate}  
\item The inverse total mean free path
  $\lambda_{\rm{tot}}^{-1}=\sum_{i=1}^{N}\lambda_{i}^{-1}$ is calculated and a distance
  $\Delta x_{1}$ to the next reaction is selected according to an exponential distribution. This is realized by using a uniformly distributed random number $0\leq r\leq 1$ via
  \begin{equation}
    \Delta x_{1} = -\lambda \ln(1-r)\,.
  \end{equation}

\item The fractional energy loss due to pair production (c.f.\ Sec.\,\ref{sct:PaP}), is limited by imposing a maximum step size $\Delta x_{2}$ given by
  \begin{equation}
    \int_{x}^{x+\Delta x_{2}}dx
      \frac{dE_{A,Z}^{e^{+}e^{-}}}{dx}(E) <\delta\,E\,,
  \end{equation}
where $\delta$ is the maximal allowed fractional energy loss. 

\item The particle is propagated over a distance 
  \begin{equation}
    \Delta x = \min(\Delta x_{1}, \Delta x_{2}, \Delta x_{3})\,,
    \label{Eq:Propagation:Algo}
  \end{equation}
where $\Delta x_{3}$ is an upper limit on the propagation step that can be provided by
the user, with typical values of $\Delta x_{3}\sim 1-50~\rm{Mpc}$. This increases the accuracy of the calculation of pair production energy losses and the accuracy of the secondary pair production spectra.

\item  If $\Delta x_{1} = \Delta x$, the particle is propagated over the path length $\Delta x_1$, where it performs an interaction. The choice of the specific interaction performed by the UHECR is taken by finding the smallest index $i$ for which
  \begin{equation}
    \sum_{a=1}^{i}\frac{\lambda_{\rm{tot}}}{\lambda_{a}}>w\,,
    \label{eq:CRPropa:SelectChannel}
  \end{equation}
for a uniformly distributed random number $0\leq w\leq 1$ . Then the continuous energy losses are applied and the algorithm is restarted from the new position and the new particles produced in the interaction are added to the list of particles to be propagated, leading to a cascade of secondary nuclei (see Fig.~\ref{fig:seccascade}). A comparison of the exclusive mean free path of the different channels $\lambda_a$ and the total mean free path for $^{47}Ca$ with the pair production loss length is shown in Fig.~\ref{CRPropa:AllReactionsInPlot}.
\begin{figure}[tbp]
\begin{center}
\includegraphics[width=\textwidth]{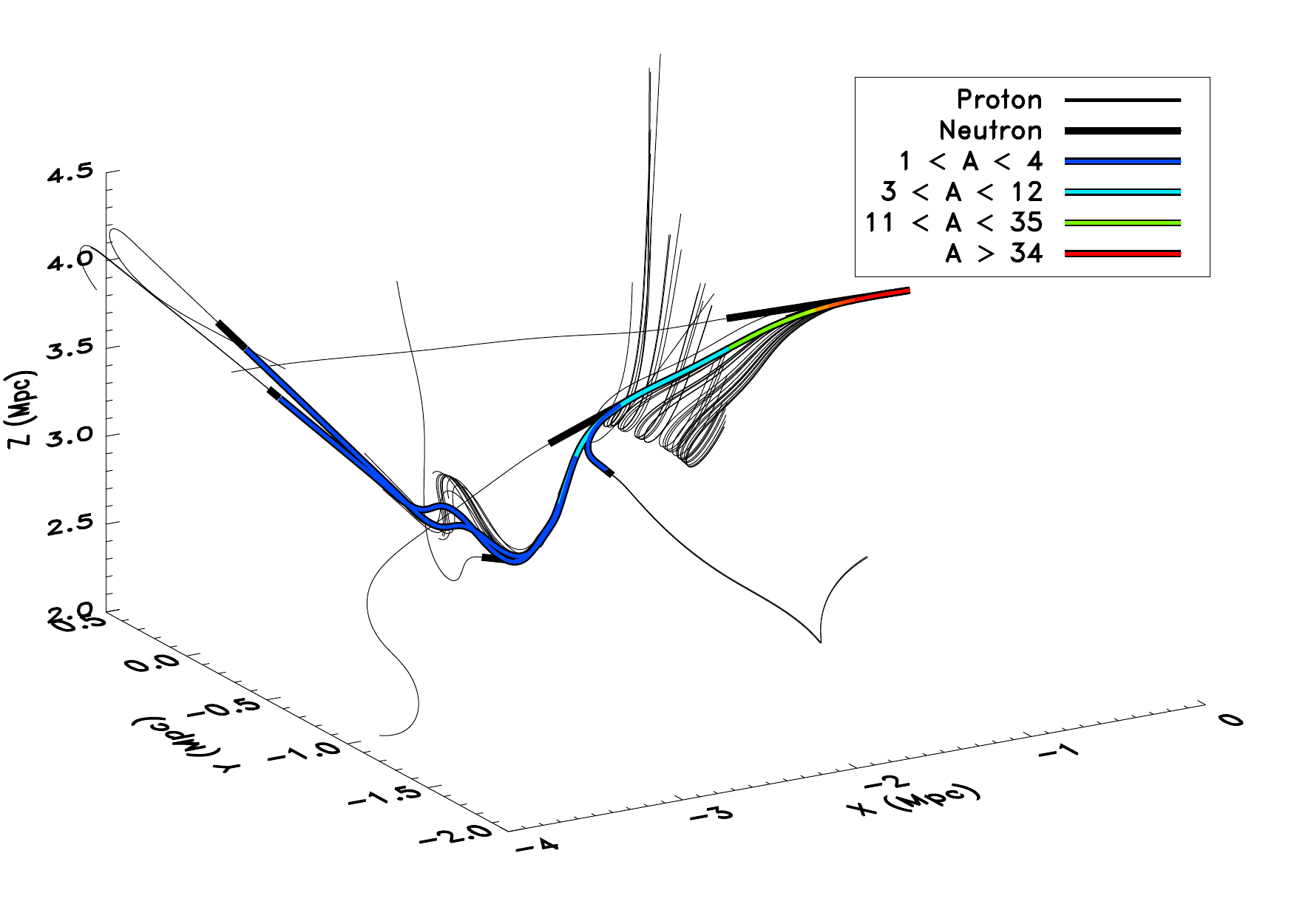}
\caption{
3D trajectory of an iron nucleus and its hadronic secondaries in the minimum of the photodisintegration mean free path ($E\sim1.2\times 10^{21}$~eV) in a high magnetic field region of a structured magnetic field ($10^{-9}{\rm G} < B < 10^{-7} {\rm G}$). The iron was injected at $x=y=z=2$~Mpc along the negative $x$ direction and enters the plotted volume on the top right. Color coded is the mass number of the secondary particles. Notice that after photodisintegration the heavy nucleus and its secondary particles have the same Lorentz factor $\Gamma\sim E/A$ and therefore secondary protons are stronger deflected than heavier nuclei due to their higher charge-to-mass ratio. 
}
\label{fig:seccascade}
\end{center}
\end{figure}

 If instead $\Delta x_{1}>\min(\Delta x_{2}, \Delta x_{3})$, then continuous energy losses are too large to allow for accurate propagation until the next interaction point or the user has requested that the maximum step size be smaller than the $\Delta x_{1}$ selected in step 1. In this case, the particle is propagated over the distance $\Delta x$ after which continuous losses are applied and the algorithm is restarted without performing any interaction.
 \end{enumerate}    

From the above description it is clear that if one of the interaction channels has a small mean free path
$\lambda_{i}$, the step size of the propagation will adjust itself automatically. 

This approach is also applied to select an exclusive channel e.g.~in case of photodisintegration: If photodisintegration is chosen to be the next reaction by the
propagation algorithm, the exclusive channel is found by applying Eq.~(\ref{eq:CRPropa:SelectChannel}). In analogy, here
$\lambda_{\rm{tot}}=(\sum_{i} \lambda_{i}^{-1})^{-1}$ is the total mean free path for
the isotope under consideration. The $\lambda_{i}$ are the mean free
path values for the exclusive channels of the corresponding isotope. 

If the user chooses to include secondary $\gamma-$rays and/or neutrinos, these neutral secondaries are propagated over a distance equal to the maximum propagation distance provided by the user minus the time of their production, such that they reach an observer after the maximum propagation time independently of the chosen 1D or 3D environment. In addition it is possible to inject a mixed nuclei composition. Since simple arguments about astrophysical acceleration mechanisms indicate that these mixed compositions should be accelerated up to a given maximum rigidity $R=E/Z$ in the sources, instead of a maximum energy, we included the option to inject up to a given maximum rigidity at the source.

The accuracy of the determination of the arrival direction and arrival time at the observer position is of course related to the actual implementation of detection and propagation in our algorithm. Besides the numerical error intrinsic to the detection algorithm, an additional error is introduced by the choice to take continuous losses into account only at the end of the time step and of course by the maximum propagation time, which can however be controlled by the user. This can be particularly relevant for bursted sources. We refer the interested reader to the manual for a deeper discussion on these issues.

\section{Example Applications\label{sct:Applications}}
In this section simulations are presented to demonstrate some features of CRPropa. All these simulations are restricted to a pure iron or a mixed galactic composition injected at the source. For the latter we adopt
Ref.~\cite{1996ApJ...465..982D}, similar to the approach of
Ref.~\cite{Allard:2005ha}. The injected power law $dN/dE\propto E^{-\alpha}$ is arbitrarily chosen to have a slope\footnote{The slope $\alpha$ of the source injection spectrum as well as other free parameters of the simulations were not optimized in order to reproduce measured UHECR data in these simple example applications.} of $\alpha=2.2$. For the 3D simulations, a (75~Mpc)$^3$ simulation cube with periodic boundary conditions is defined and filled with the large scale structure extragalactic magnetic fields (LSS-EGMF) from the cosmological simulations given in Ref.~\cite{2004NuPhS.136..224S}.

\subsection{Completeness of the Photodisintegration Cross Section Tables}
\begin{figure}[tbp]
  \centering
      \includegraphics[width=1.\textwidth]
                      {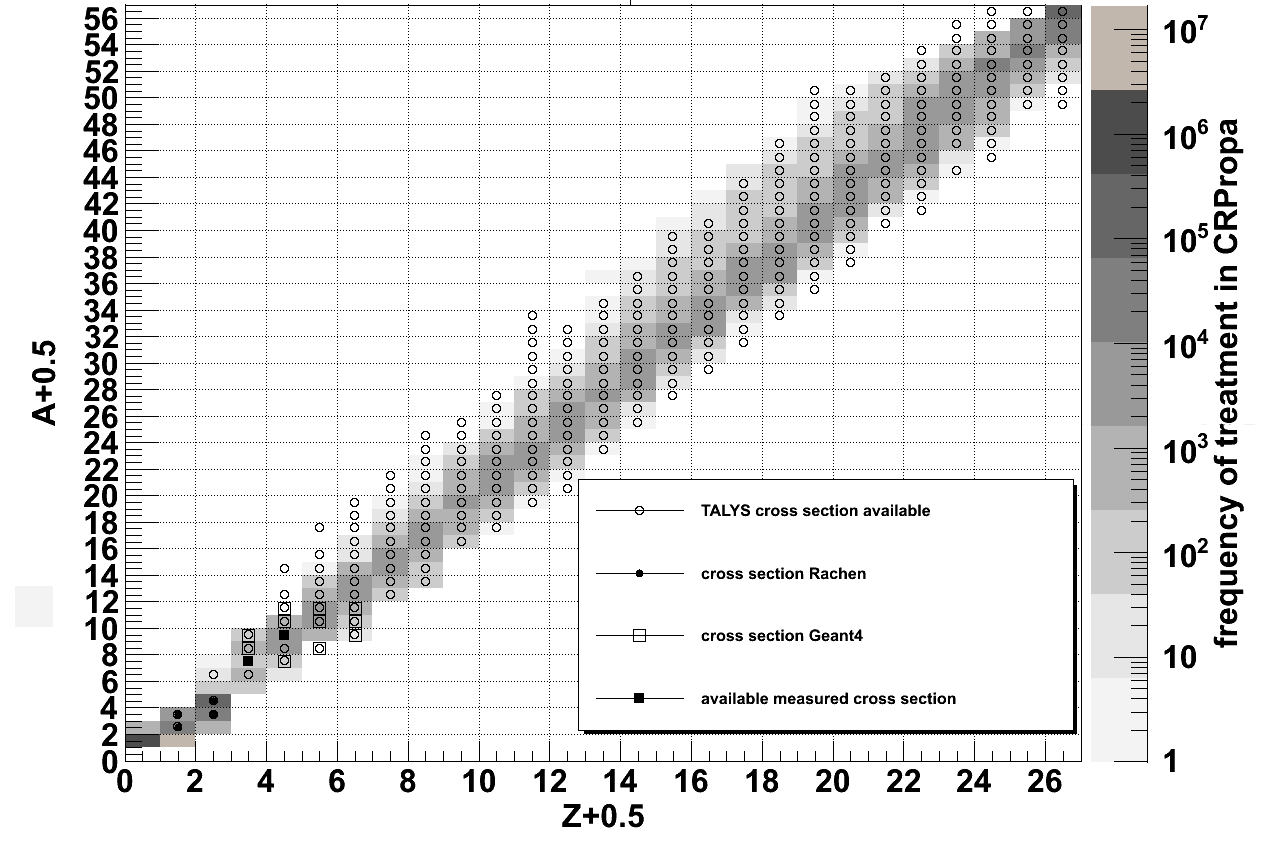}
    \caption{Isotopes and their frequency of occurrence during the propagation simulated by CRPropa for $10^{5}$ injected iron nuclei rectilinearly propagated over a distance up to 1000~Mpc. Note that this does not reflect an observable abundance, which would also depend on the time of survival of the produced isotope.
      \label{CRPropa:UsedNuclei:All}
    }
\end{figure}
To verify the completeness of the implemented photonuclear cross section tables presented in Sec.\,\ref{sct:CRPropa:ThePDXS}, i.e.\ to verify whether cross sections are available for all nuclei that occur during propagation, 1D simulations have been performed.
To this purpose, $10^{5}$ iron nuclei were injected with a $dN/dE\propto E^{-1}$ spectrum in the energy range $1\,{\rm EeV}\leq E\leq 56\times10^3\,$EeV from a uniform source distribution extending up to a distance of 1~Gpc from the observer. In this simulation, all particles that were created and propagated within CRPropa were recorded in a two dimensional histogram (see Fig.~\ref{CRPropa:UsedNuclei:All}) displaying the frequency of occurrence of isotopes in the simulation as a function of their mass and atomic number $A,Z$. In this figure, symbols are given to mark the isotopes for which cross sections are available in CRPropa and the type of the marker identifies the source of the cross section as listed in Sec.\,\ref{sct:CRPropa:ThePDXS}. At the upper right part of Fig.~\ref{CRPropa:UsedNuclei:All} some nuclei are created but do not have a photodisintegration cross section assigned. This is not problematic since all of them are far off the valley of stability and can be expected to decay very quickly. A closer look at the remaining light nuclei suggests that five of them  cannot be handled due to missing cross section data: $^5$He, $^5$Li, $^9$B, $^7$He and $^6$Be. However, all these nuclei have a half-life time smaller than $10^{-15}$~s and, hence, are too short-lived to undergo photodisintegration. Finally,
due to mass and charge loss by pion production, ``nuclei'' which consist only of neutrons or protons can be observed in figure \ref{CRPropa:UsedNuclei:All}. This is a purely technical artifact and those nuclei will immediately decay in CRPropa. 
Thus, this basic example simulation suggests that the compilation of photonuclear cross sections used in CRPropa\,2.0 is complete for an application in UHECR astrophysics.

\subsection{Comparison of the Photodisintegration Cross Section Tables with PSB\label{sct:Applications:PSB}}
\begin{figure}
  \centering
  \subfigure[primary cosmic rays]{
    \includegraphics[width=.47\textwidth]{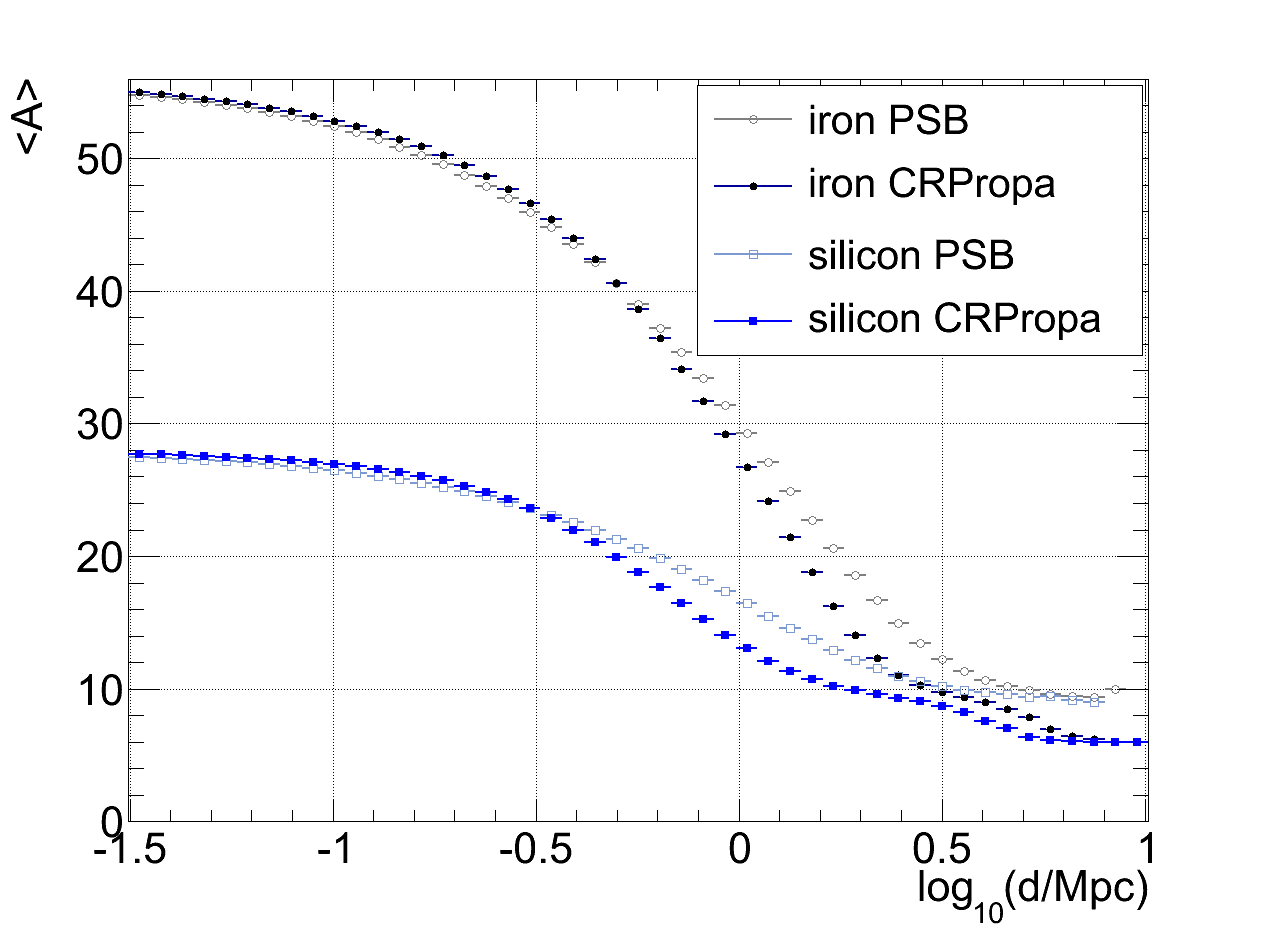}
    \label{Plots:PSBcomparison:primary}
  }
  \subfigure[cosmic rays including secondaries]{
    \includegraphics[width=.47\textwidth]{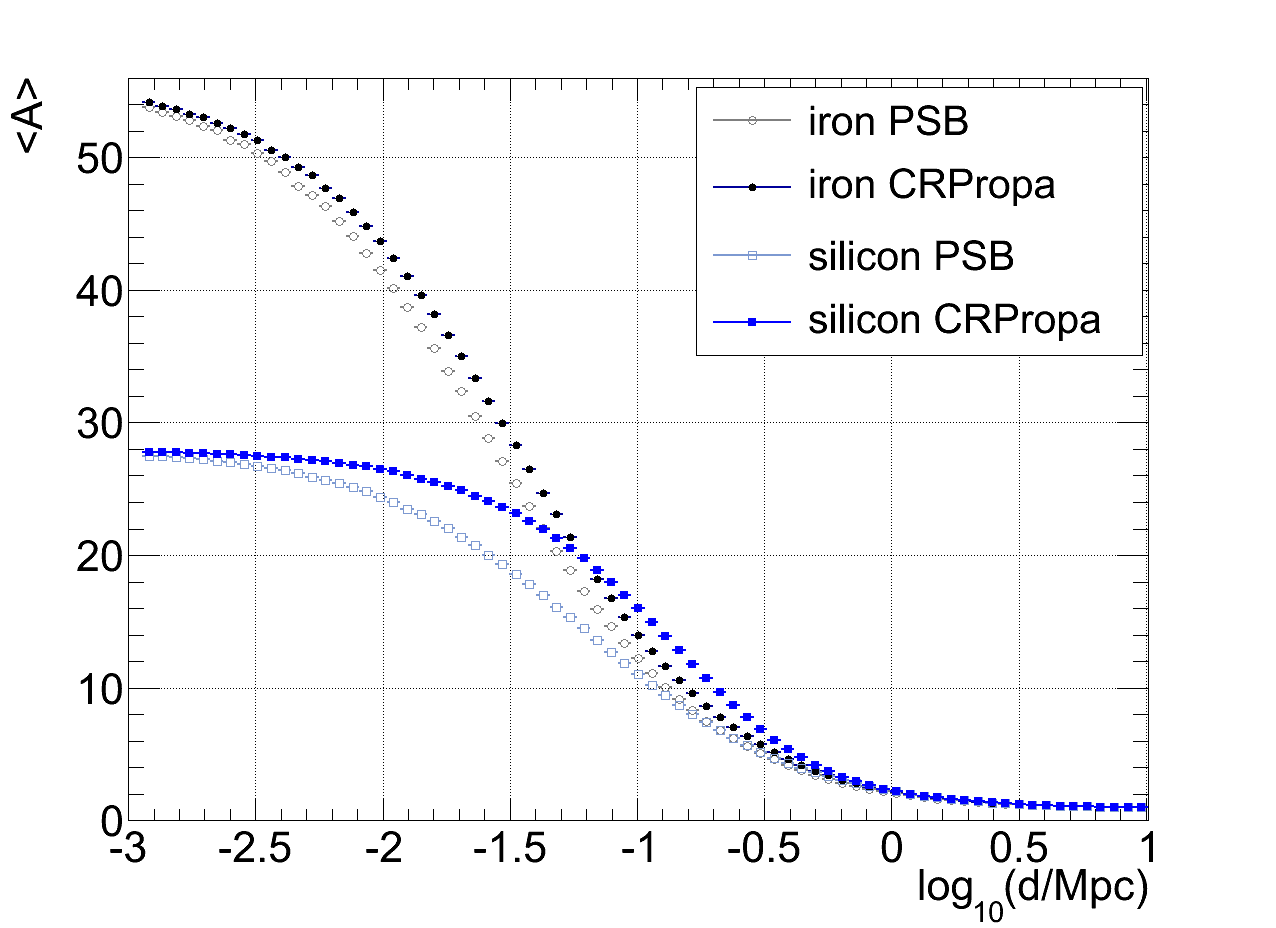}
    \label{Plots:PSBcomparison:secondaries}
  }
  \caption{
    Average nuclear mass $\langle A \rangle$ as a function of the distance from the source, resulting from the CRPropa photodisintegration cross section tables (see Sec.\,\ref{sct:CRPropa:ThePDXS}) (solid markers) and the PSB cross section tables \cite{Puget:1976nz} (open markers). The 1D simulations assume emission of pure iron (black circles) and pure silicon (blue rectangles) with an injection rigidity of $R=38.4\,$EeV. 
    \newline(a) Average mass number of the primary cosmic rays, disregarding all nuclei with $A<5$.
    \newline(b) Average mass number of all cosmic rays, including secondaries.
    \label{Plots:PSBcomparison}}
\end{figure}
Figure~\ref{Plots:PSBcomparison} depicts example 1D simulations comparing TALYS (Sec.\,\ref{sct:CRPropa:ThePDXS}) and PSB cross sections \cite{Puget:1976nz}. 
These simulations show the average mass number $\langle A \rangle$ of UHECRs as a function of the distance from the source for the two cross section models under consideration. Here two different  cases, namely pure iron and pure silicon injection, with the same rigidity at injection $R=38.4\,$EeV (corresponding to injection energies of $E=$1000 and 538$\,$EeV respectively), are considered. Furthermore these two injection cases have been provided both for a scenario excluding secondary particles (a) and including secondary particles (b). In these simulations, pair production, pion production and redshift evolution have been disabled, photodisintegration is considered on the CMB only and the cosmic rays are tracked as long as their energy is above 0.1\,EeV. In the PSB case, decay has been disabled in order to strictly follow the reduced reaction network of \cite{Puget:1976nz}. As the PSB tables do not provide photodisintegration cross sections for $5\leq A \leq 8$, particles that end up in this region are set to immediately photodisintegrate to $A=4$ plus secondaries. 

In scenario (a), in order to consider only the primary cosmic rays, all nuclei with $A<5$ have been ignored. This case can be compared with, for instance, Fig. 5 of \cite{2005APh....23..191K}. 

As noted in \cite{2005APh....23..191K}, the PSB agreement with experimental data is not as good as the one obtained with the Lorentzian parameterization of TALYS. Moreover, the PSB case employs a reduced reaction network involving only one nucleus for each atomic mass number $A$ up to $^{56}$Fe, whereas 287 nuclides with their photodisintegration cross sections have been implemented in CRPropa. Figure\,\ref{Plots:PSBcomparison}  shows the effect of these differences on the average mass of the primary particle. In particular it can be seen that CRPRopa employing TALYS (plus low-mass extensions) results on average in a faster photodisintegration rate than PSB does. 

In scenario (b) all photodisintegrated particles are included. Due to the light secondary particles that are produced with each photodisintegration of the primary particle, the average mass number decreases faster in this scenario than in scenario (a). This shows the importance of taking secondary particles into account when predicting the average mass number.

A noticeable feature in this scenario is that, for CRPropa, after a certain propagation length the average mass number of injected silicon exceeds the average mass number of injected iron. This is due to the larger number of light secondaries disintegrated off the iron nucleus. This, in combination with the cross section dependence on the mass number, can cause a lower total {\em average\/} mass at a certain distance, even though the primary cosmic ray still has a higher mass on average.
 
Furthermore, in this scenario both iron and silicon injection show, at all distances, an average mass for the CRPropa tables larger or equal to the average mass for the PSB tables. This can be tracked to a difference in the type of secondaries that are created. In CRPropa photodisintegration can yield secondaries of mass number up to four ($n$, $p$, $d$, $t$, $^3$He and $^4$He). In the PSB case all secondaries, with the exception of the reaction $\gamma + ^9\!{\rm Be} \rightarrow 2\alpha + n$, have a mass number of one, therefore decreasing the average mass number with respect to the CRPropa case.

\subsection{1D: Influence of Chemical Composition and Cosmological Evolution}
\begin{figure}
  \centering
  \subfigure[galactic composition]{
    \includegraphics[width=.47\textwidth]{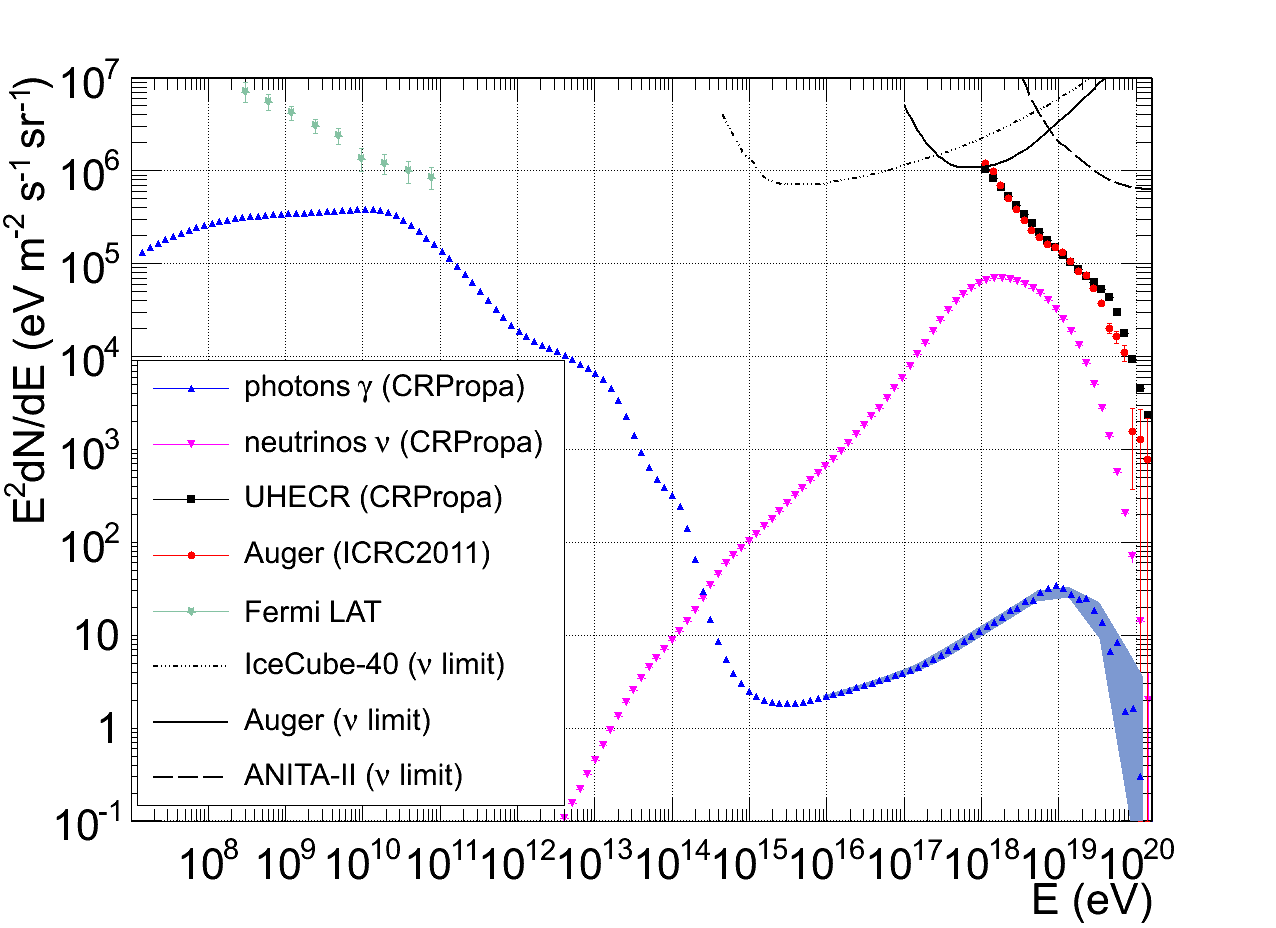}
    \label{Plots:1D_1st_Example_top_gal}
  }
  \subfigure[pure iron]{
    \includegraphics[width=.47\textwidth]{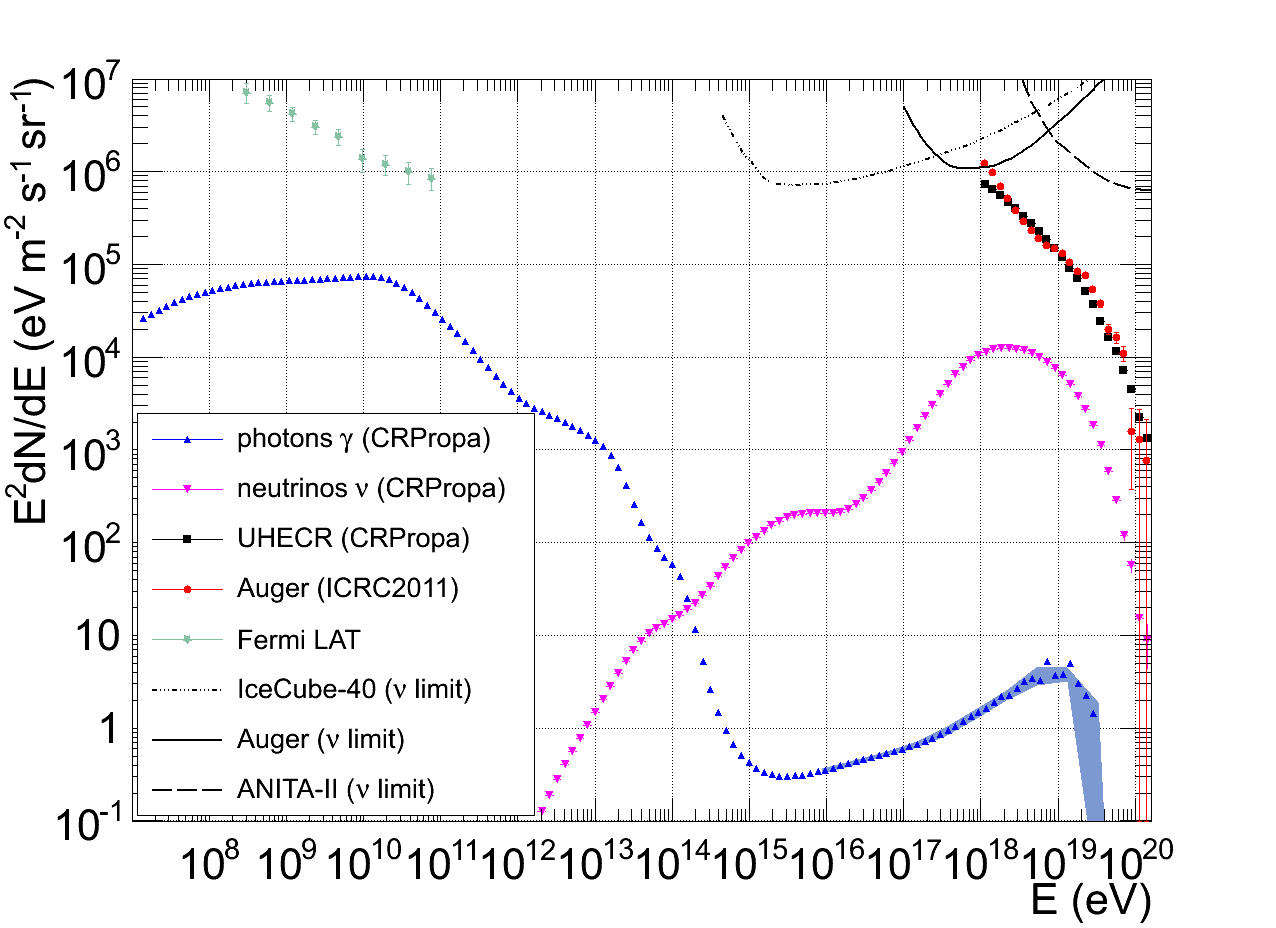}
    \label{Plots:1D_1st_Example_top_iron}
  }

  \subfigure[Abundance ($E>1~\rm EeV$).]{
    \includegraphics[width=.47\textwidth]{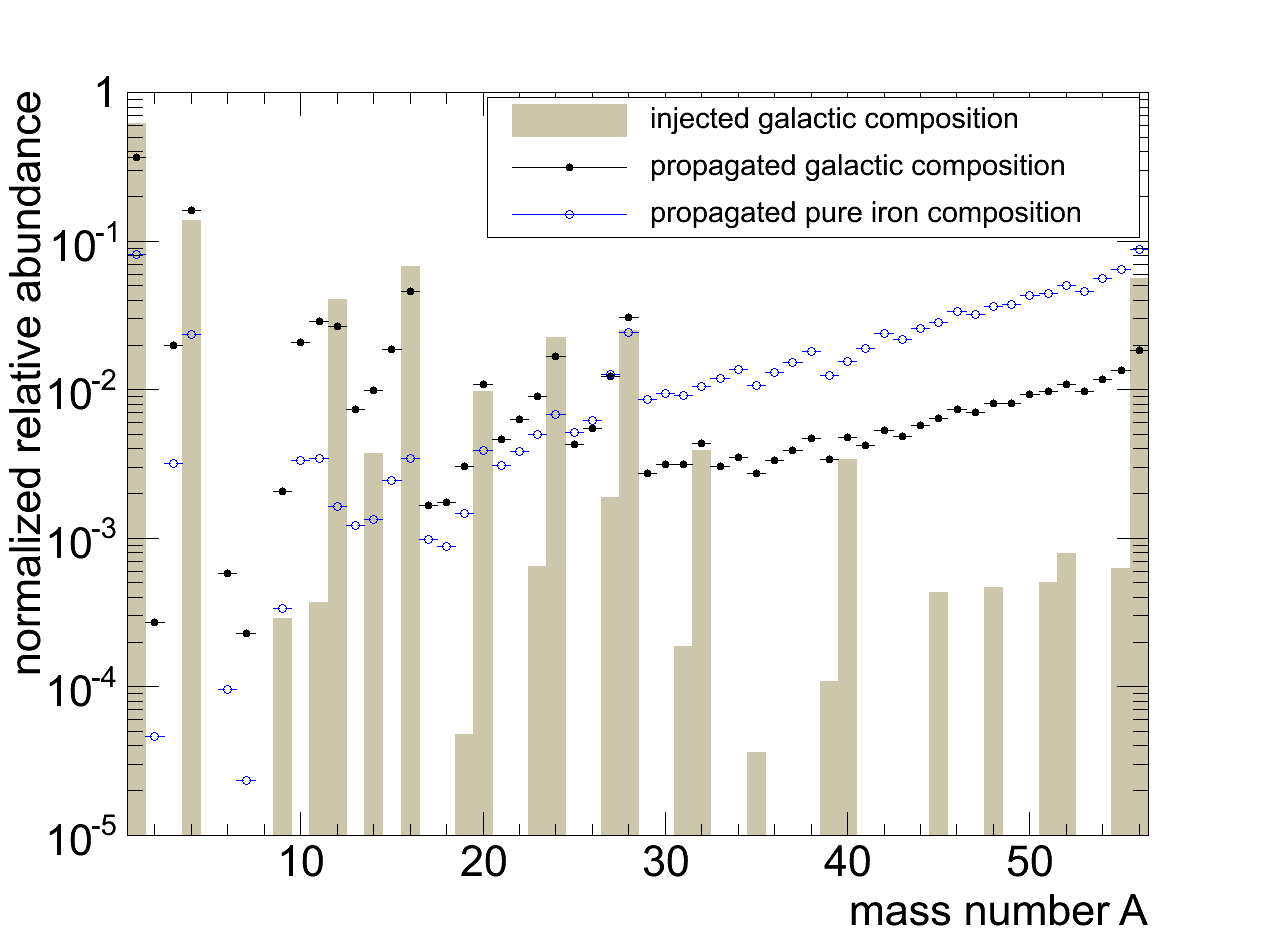}
    \label{Plots:1D_1st_Example_bottom}
  }
  \caption{
    Results of a 1D simulation with CRPropa\,2.0 taking into account the cosmological expansion and a comoving source evolution scaling as $(1+z)^4$ up to $z_{\rm max}=2$.
     \newline(Upper panels:) The simulated UHECR flux (black rectangles) has been normalized to the Pierre Auger spectrum (red dots)~\cite{Abraham:2010mj,Abreu:2011pj}. The spectra of secondary $\gamma-$rays (blue triangles) and neutrinos (magenta triangles) have been normalized accordingly. The neutrino flux shown is the single-flavor flux, assuming a ratio of 1:1:1. This flux can be compared with the single-flavor neutrino limits (black lines) \cite{Abbasi:2011ji,Abreu:2011zze,Gorham:2010kv}.   Green stars show the isotropic $\gamma$-ray flux measured by Fermi-LAT \cite{Abdo:2010nz}. In the left upper panel a galactic mixed composition has been injected at the source while in the right upper panel a pure iron composition has been injected.
    \newline (Lower panel:) Abundance of UHE nuclei above 1 EeV after propagation in case of a pure iron (blue open circles) and mixed galactic composition (black solid circles) injected. For comparison, the original galactic composition (light brown area) is also shown.
    \label{Plots:1D_1st_Example}}
\end{figure}
The advantage of the 1D mode in CRPropa is that one can include the cosmological as well as the source evolution as function of the distance to the observer. We demonstrate this by using two simulations which only differ in the composition injected at the sources, namely pure iron injection or a galactic mixed composition~\cite{1996ApJ...465..982D,Allard:2005ha}. This allows one to investigate the influence of the poorly known initial composition. The parameters of these simulations are as follows: UHECRs are injected with an $E^{-2.2}$ spectrum up to a rigidity of $R= 384.6\,$EeV from a continuous source distribution with comoving injection rate scaling as $(1+z)^4$ up to $z_{\rm max}=2$. The cosmological evolution is characterized by a concordance $\Lambda$CDM Universe with a cosmological constant ($\Omega_{m}=0.3$, $\Omega_{\lambda}=0.7$) using a Hubble constant of $H_{0}=72\rm \;km\;s^{-1}\;Mpc^{-1}$. 

The results are shown in Fig.~\ref{Plots:1D_1st_Example}. The simulated UHECR flux is normalized to the spectrum observed by the Pierre Auger Observatory at an energy of $10^{19}~\rm EeV$ \cite{Abraham:2010mj,Abreu:2011pj} and the secondary $\gamma-$ray and neutrino fluxes are normalized accordingly. The reduction of the flux of observable secondary neutrinos and $\gamma-$rays for pure iron injection shown in Fig.~\ref{Plots:1D_1st_Example_top_iron} relative to Fig.~\ref{Plots:1D_1st_Example_top_gal} for a mixed injection composition is due to photodisintegration dominating with respect to photopion production for the heavier composition. The resulting simulated abundance of UHE nuclei for $E>1\,$EeV is shown in Fig~\ref{Plots:1D_1st_Example_bottom}. To compare with the propagated composition, the injected galactic composition is also shown.

Note that photons from nuclear de-excitation during a photodisintegration event are currently not taken into account in CRPropa. This might cause a moderate increase of the flux of photons at energies below $\sim 10^{17}\,$eV as discussed in e.g.~\cite{Anchordoqui:2006pd,Murase:2010va}.

\subsection{3D: Continuous Source Distributions following the Large Scale Structure\label{3DEvents}}
\begin{figure}
  \centering
  \subfigure[energy spectra]{
     \includegraphics[width=0.475\textwidth]{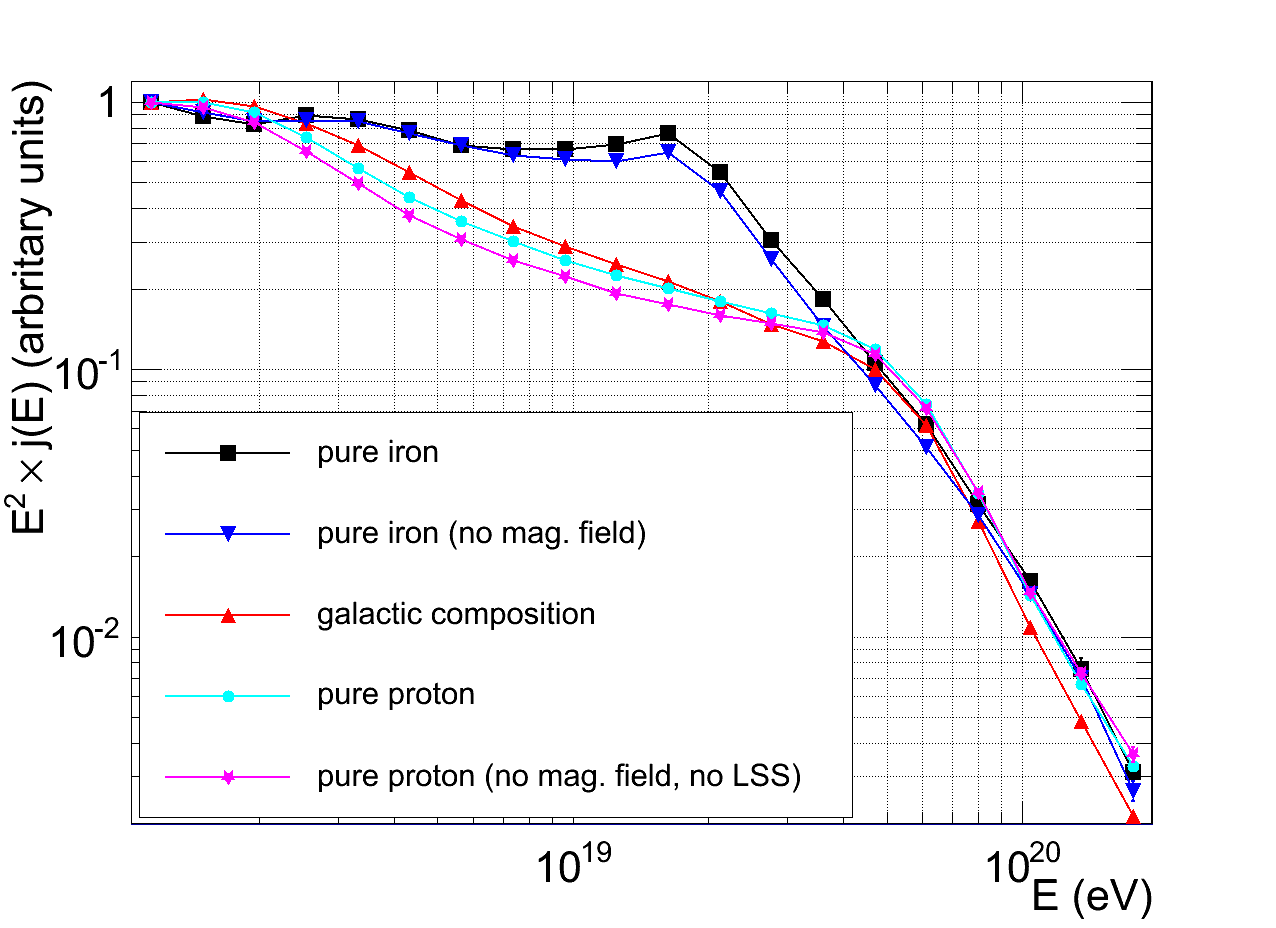}
     \label{Plots:3D_1st_Example:Event:Spectra} 
   }
   \subfigure[mass spectra]{
     \includegraphics[width=0.475\textwidth]{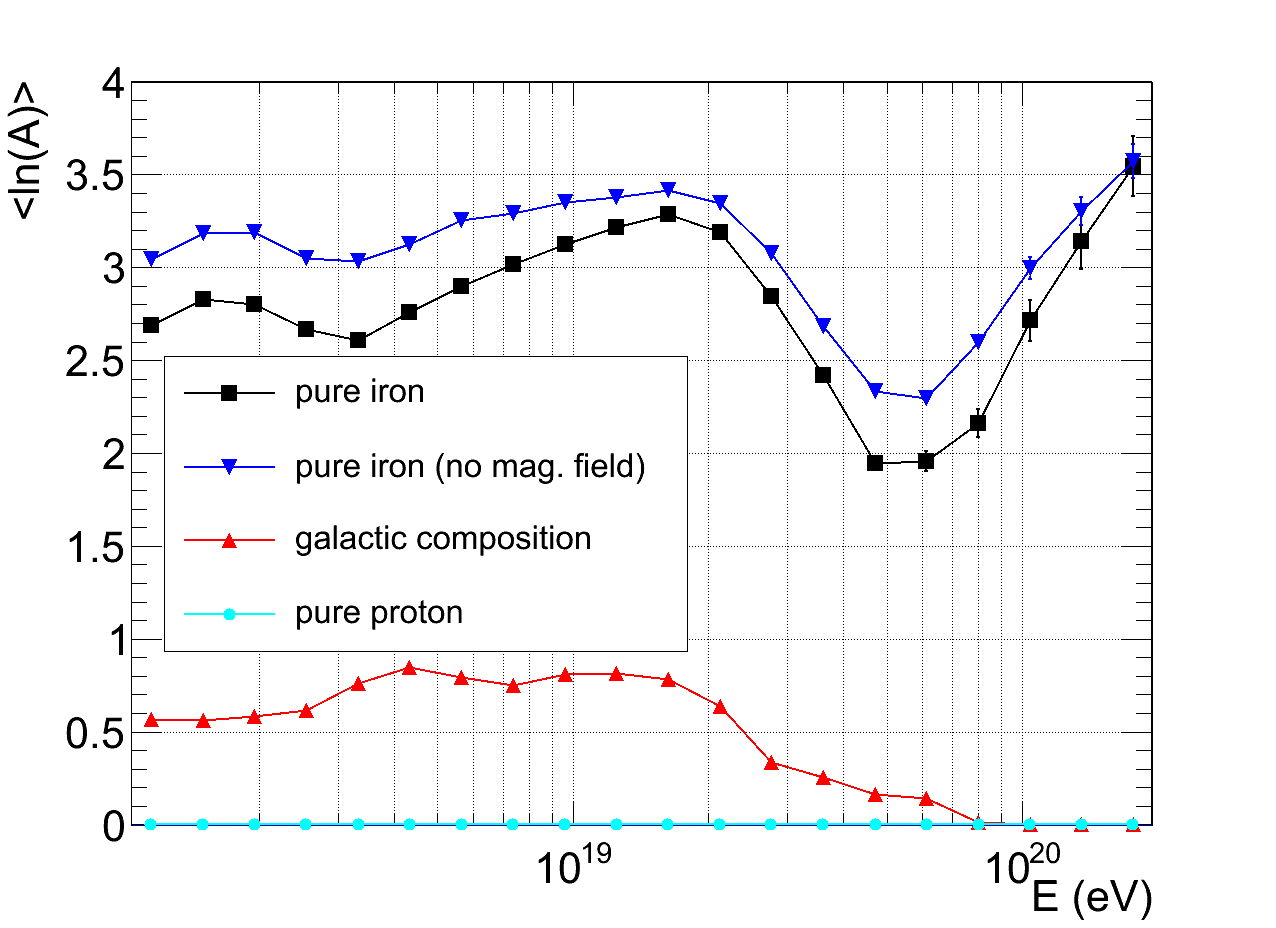}
     \label{Plots:3D_1st_Example:Event:Compo}
   }
  \subfigure[source distance vs travel time]{
    \includegraphics[width=0.475\textwidth]{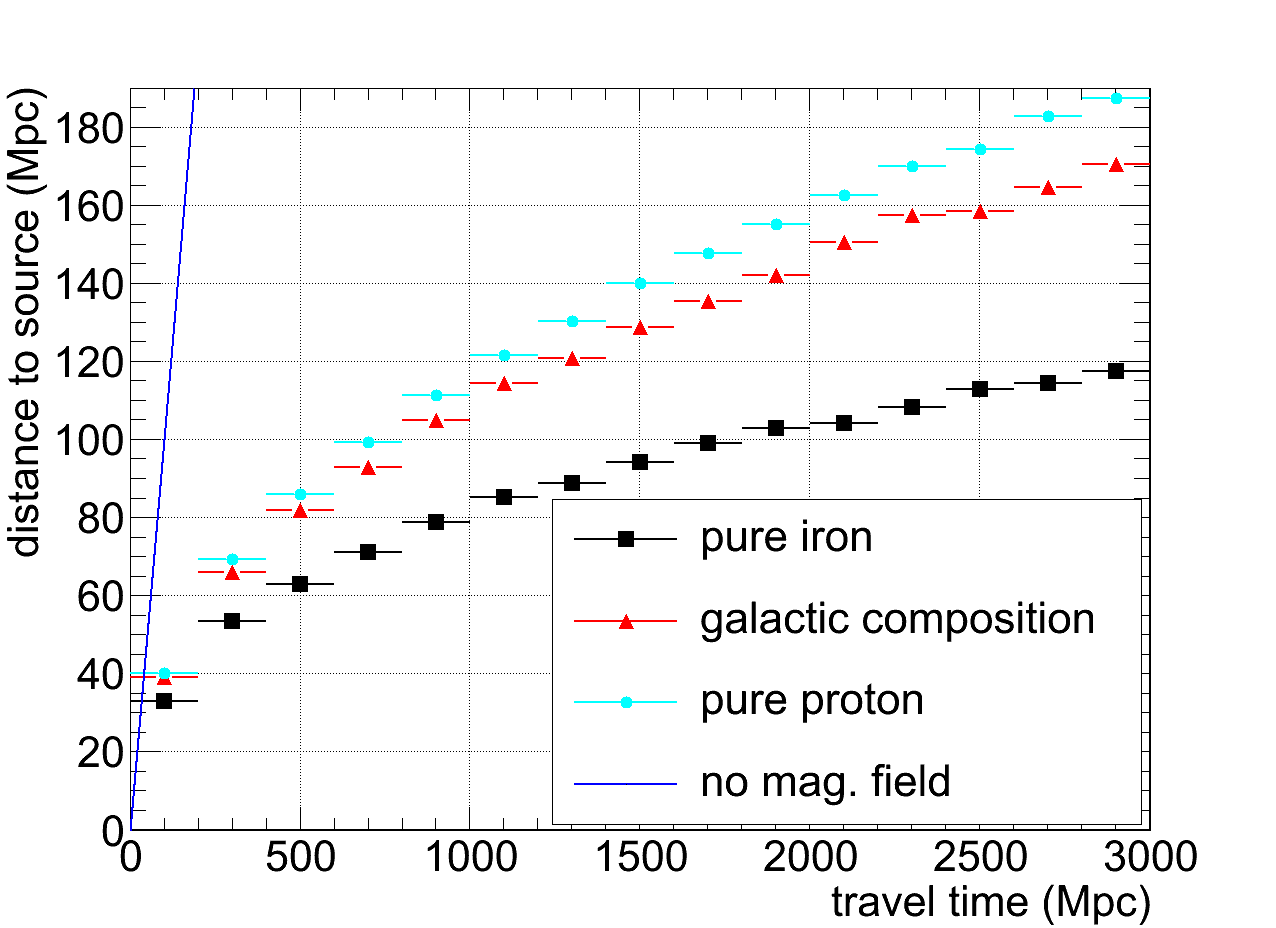}
    \label{Plots:3D_1st_Example:Event:DistVsTime}
  }
  \caption{
Example simulations of a continuous source distribution which follows the LSS density in the scenario of Ref.~\cite{2004NuPhS.136..224S}. Magnetic deflections of the UHECRs in the corresponding LSS magnetic fields are taken into account in this 3D simulation. A pure iron composition (black rectangles), a galactic composition (red triangles) and a pure proton composition (light blue dots) have been injected at the source.
\newline (a), (b) UHECR flux and average mass number $\langle A \rangle$ as function of energy. Apart from the three cases for injected composition discussed in the text, which are shown including deflections in the LSS-EGMF, a pure iron simulation without deflection is shown for comparison (dark blue triangles). In the case of the UHECR flux a simulation with pure proton injection without deflections and with a flat continuous source distribution (not following the LSS density) has been added for comparison as well (magenta stars). The flux is normalized to unity in the first bin to allow for a better comparison of the spectral shape. 
\newline (c) Distance of the UHECRs from their source as function of the propagation time for all cosmic rays above 1 EeV, including a scenario without deflections (dark blue line).
\label{Plots:3D_1st_Example:Event}
} 
\end{figure}
A very attractive feature of CRPropa is the possibility to study the effect of the presence of a large scale structure extragalactic magnetic field (LSS-EGMF) on the UHECR spectrum, composition and anisotropy in 3D simulations. In the following example we study how a continuous source distribution following the LSS baryon density and deflections in the corresponding LSS-EGMF in the scenario of Ref.~\cite{2004NuPhS.136..224S} can influence these quantities. To this end we inject a power law of $dN/dE \propto E^{-1}$ which we then reweigh to a power law of $dN/dE \propto E^{-2.2}$, a well known trick to achieve sufficient statistics at high energy. As in the previous example, we consider a pure iron and a mixed galactic-like composition at injection and for comparison a pure proton composition has been added as well. Particles are injected up to a rigidity of $R=384.6\,$EeV and are tracked as long as their energy is above 1~EeV. The detection occurs on a sphere centered around the observer, called ``sphere around observer'' in the code, with a radius of $\simeq1\,$Mpc. The observer is placed in a magnetic environment that is similar to what is found in the vicinity of our Galaxy. 

Fig.~\ref{Plots:3D_1st_Example:Event:Spectra} shows that, in case of a pure iron injection, a bump in the UHECR spectrum is predicted at $\sim15\,$EeV, which does not occur if a mixed galactic composition or pure proton composition is injected. Furthermore, the simulated spectra are not strongly affected by the presence of the LSS magnetic field or LSS source density.  In contrast, the propagated composition of the pure iron injection case is affected by deflections, as shown in Fig.~\ref{Plots:3D_1st_Example:Event:Compo}, since deflections increase the propagation path length, thereby enhancing interactions and reducing the average mass number $\langle A \rangle$ at detection. Finally, Fig.~\ref{Plots:3D_1st_Example:Event:DistVsTime} illustrates that the pure iron case shows a smaller horizon for energies $E\gtrsim1\,$EeV when compared to the case of an injected galactic or pure proton composition. This is mostly due to increased deflections and thus more interactions in case of primary iron nuclei.

\subsection{3D: Simulating Observables at a given Distance from a Source}
Apart from detecting particles on spheres around the observer, cf.~Sec.\,\ref{3DEvents}, CRPropa also allows one to detect particles on spheres around sources. In this detection mode a simulated UHECR trajectory crossing a given sphere in a 3D simulation will be written to the output file. This allows one to study e.g.\ spectrum, composition and anisotropy of the UHECR flux from a given source as function of the distance, corresponding to the radii of the spheres.

\begin{figure}
  \centering
  \subfigure[energy spectra]{
    \includegraphics[width=0.55\textwidth]{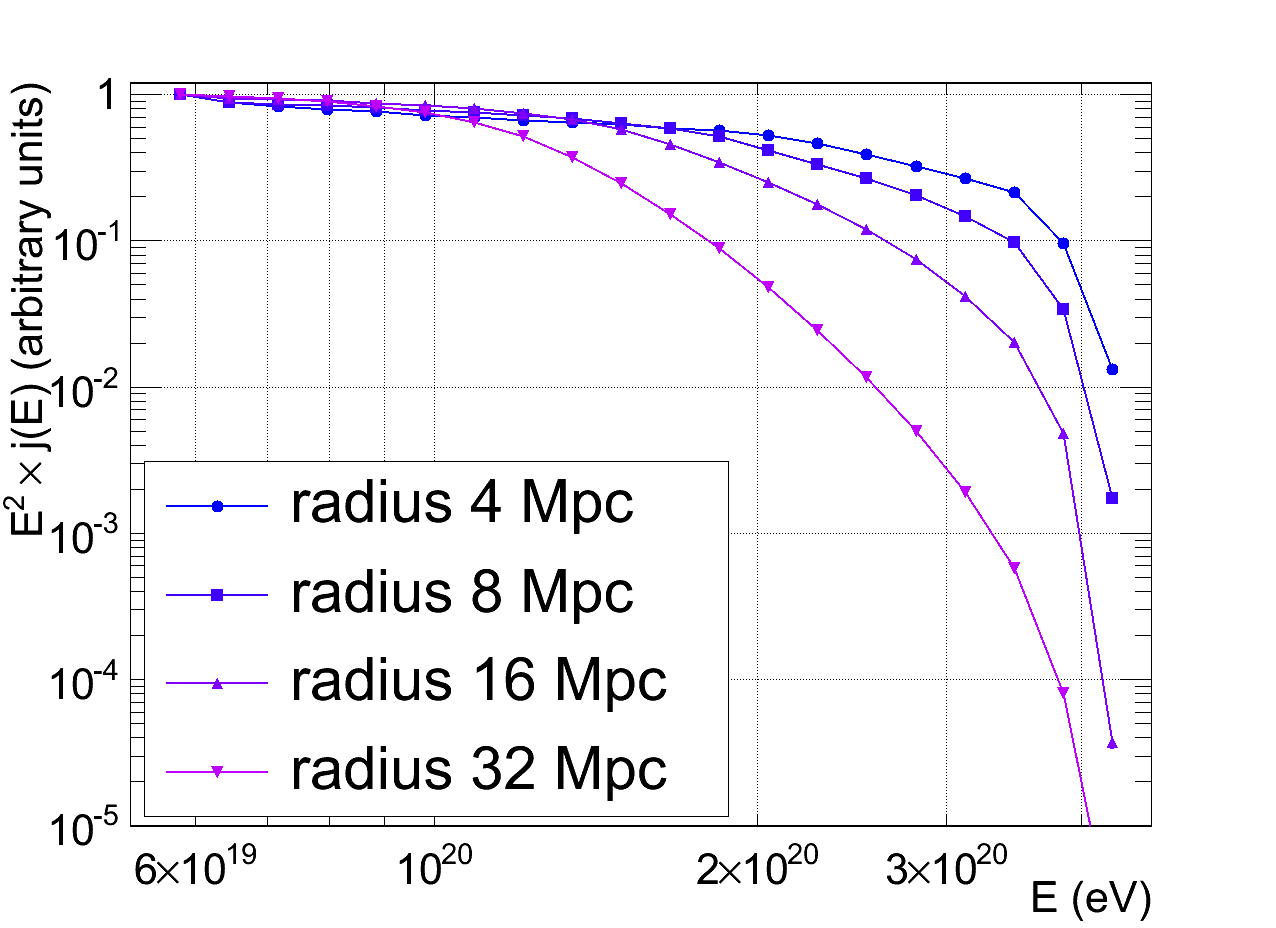}
    \label{Plots:3D_1st_Example:SpheresAroundSources:Spec}
  }
  \subfigure[sky maps]{
    \includegraphics[width=1.\textwidth]{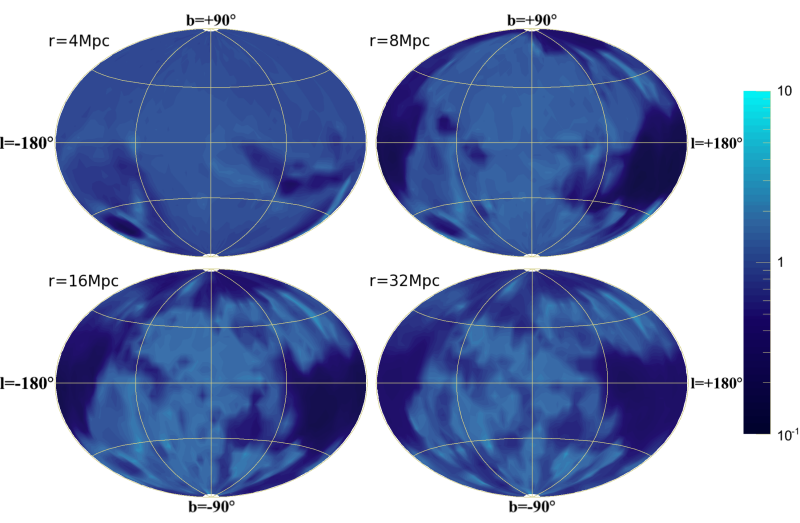}
    \label{Plots:3D_1st_Example:SpheresAroundSources:combined}
  } 

  \caption{Spectrum and sky maps of UHECR above $55\,$EeV after propagating in a $(75~\rm Mpc)^3$ simulation cube filled with an LSS-EGMF in the scenario of Ref.~\cite{2004NuPhS.136..224S}. In this simulation detection spheres have been placed around the single source present in the simulated environment with the four radii 4, 8, 16, and 32~Mpc and the injection spectrum has been reweighted to a $dN/dE \propto E^{-2.2}$ source spectrum.
\newline (a) Differential spectrum $dN/dE$ multiplied by $E^{2}$ as detected at the four spheres.
\newline (b) Hammer-Aitoff projections of the arrival directions of the simulated UHECR trajectories as registered on the different detection spheres around the source. 
\label{Plots:3D_1st_Example:SpheresAroundSources}
} 
\end{figure}

As an example a galactic composition is injected from the center of a (75~Mpc)$^3$ simulation box filled with the magnetic field configuration as given in the scenario of Ref.~\cite{2004NuPhS.136..224S}. Detection spheres with radii of 4, 8, 16 and 32~Mpc are placed around the source. The initial $dN/dE \propto E^{-1}$ spectrum up to a rigidity of $R=384.6~\rm EeV$ is reweighted to a  $dN/dE \propto E^{-2.2}$ spectrum. Only particles which have an energy larger than $55\,$EeV are taken into account. As shown in Fig.~\ref{Plots:3D_1st_Example:SpheresAroundSources}, the flux of cosmic rays at the highest energies is suppressed by particle interactions as the distance from the source increases. Furthermore, the cosmic ray distribution becomes more anisotropic with increasing distance from the source due to increasing deflections in the large scale magnetic field structure. This latter effect is exemplified in the sky maps shown in Fig.~\ref{Plots:3D_1st_Example:SpheresAroundSources:combined}.  

\section{Summary and Outlook}
In the present paper we have presented the new version of our UHECR propagation code CRPropa, a numerical tool to study the effect of extragalactic propagation on the spectrum, chemical composition and distribution of arrival directions of UHECRs on Earth. The main new feature introduced in this new version 2.0 is the propagation of UHE nuclei, taking also into account their interactions with the IGM, in particular photodisintegration, which is modeled according to the numerical framework TALYS. 
As photodisintegration introduced many more interaction channels than were present in the previous version of CRPropa, we needed to substantially improve the propagation algorithm both in efficiency and in accuracy. We also updated the default model for the extragalactic infrared light to a more recent one. 
CRPropa\,2.0 can now be used to compute the main observable quantities related to UHECR propagation with the accuracy required by present data: particle spectra, mass composition and arrival direction on Earth, for highly customizable realizations of the IGM, including source distributions and magnetic fields. In addition, the spectra of secondary neutrinos and electromagnetic cascades can be computed down to MeV energies.
One of the extensions planned for the future is to include a module to take into account the effect of galactic magnetic fields on deflections.

\section{Acknowledgements}
We warmly thank Ricard Tom\`{a}s and Mariam T\'{o}rtola for many interesting discussions and help with several parts of the code. 
This work was supported by the Deutsche Forschungsgemeinschaft through the collaborative research centre SFB 676, by BMBF under grants 05A11GU1 and 05A11PX1, and by the ``Helmholtz Alliance for Astroparticle Phyics (HAP)'' funded by the Initiative and Networking Fund of the Helmholtz Association. GS acknowledges support from the State of Hamburg, through the Collaborative Research program ``Connecting Particles with the Cosmos''. LM acknowledges support from the Alexander von Humboldt foundation.

\bibliography{CRPpaper}
\bibliographystyle{model1-num-names}

\end{document}